\documentclass[preprint,12pt]{elsarticle}

\usepackage{amssymb}
\usepackage{graphicx}
\usepackage{url}

\usepackage{algorithm}
\usepackage{algorithmic}
\usepackage{mathtools}
\usepackage{color}
\usepackage{listings}
\usepackage{amsfonts}
\usepackage{amsmath}
\usepackage{array}
\newtheorem{theorem}{Definition}
\newcommand{\Act}{\text{TL}}    

\newcommand{\Rt}{\mathbb{R}}
\newdefinition{rmk}{Remark}

\begin{document}

\begin{frontmatter}
%
%
%
\title{Dependability Modeling and Optimization of Triple Modular Redundancy Partitioning for SRAM-based FPGAs}

\author[label1]{Khaza Anuarul Hoque\corref{mycorrespondingauthor}}
\cortext[mycorrespondingauthor]{Corresponding author}
\ead{hoquek@missouri.edu}
\author[label3]{Otmane Ait Mohamed}
\author[label3]{Yvon Savaria}

\address[label1]{University of Missouri, Columbia, USA}
\address[label2]{Concordia University, Montreal, Canada}
\address[label3]{Polytechnique Montr\'eal, Montreal, Canada}

\begin{abstract}
SRAM-based FPGAs are popular in the aerospace industry for their field programmability and low cost. However, they suffer from cosmic radiation-induced Single Event Upsets (SEUs). Triple Modular Redundancy (TMR) is a well-known technique to mitigate SEUs in FPGAs that is often used with another SEU mitigation technique known as configuration scrubbing. Traditional TMR provides protection against a single fault at a time, while partitioned TMR provides improved reliability and availability. In this paper, we present a methodology to analyze TMR partitioning at early design stage using probabilistic model checking. The proposed formal model can capture both single and multiple-cell upset scenarios, regardless of any assumption of equal partition sizes. Starting with a high-level description of a design, a Markov model is constructed from the Data Flow Graph (DFG) using a specified number of partitions, a component characterization library and a user defined scrub rate. Such a model and exhaustive analysis captures all the considered failures and repairs possible in the system within the radiation environment. Various reliability and availability properties are then verified automatically using the PRISM model checker exploring the relationship between the scrub frequency and the number of TMR partitions required to meet the design requirements. Also, the reported results show that based on a known voter failure rate, it is possible to find an optimal number of partitions at early design stages using our proposed method.

\end{abstract}

\end{frontmatter}
\section{Introduction}
Field programmability, low manufacturing cost, and other advantages make SRAM-based FPGAs an attractive option compared to ASICs for space applications. Unfortunately, the main disadvantage of these devices is their sensitivity to cosmic radiation effects commonly known as Single Event Upsets (SEUs) \cite{XilinxRosetta}. SEUs occur when one or more bits in configuration memory change state due to a radiation event. If only one bit in the configuration memory is affected, then it is called a Single-Bit Upset (SBU). A Single-Cell Upset (SCU) is defined as an event that induces a single-bit upset at a time. If more than one bit is affected at a time in multiple storage locations, the event is known as a Multiple-Cell Upset (MCU). Since the FPGA configuration bits are stored in volatile SRAMs, SEUs are a major concern for the successful operation of safety-critical systems. To deal with SEUs, designers mostly rely on redundancy-based solutions, such as Triple Modular Redundancy (TMR) \cite{XilinxTMR} and \emph{configuration memory (Configuration Bits) scrubbing} \cite{NASAScrub}. TMR is a well-known technique for fault mitigation in which three redundant copies of the same logic perform the same task in parallel. A majority voter chooses the correct output from these three copies. Scrubbing uses a background task that corrects the SEUs using error-correcting code memory or a redundant copy of data, either periodically or on detection of an SEU.


Reliability analysis of TMR and their related improvements have been studied for a long time and widely reported in the literature. By contrast, partitioning of TMR for reliability improvement got less attention from the research community. In \cite{Related-2TMR}, the authors present Markov models for partitioned TMR and showed how TMR partitioning can help improving the reliability. In their models, the TMR partitions were assumed to be equal in size. This assumption reduces the complexity of the model. However, this is a clear limitation, because in a real design, it is not always possible or even desired to have partitions of equal size. An optimal partitioning for TMR logic was reported in \cite{Related-3TMRoptimal} and the relationship between DCEs (Domain Crossing Events) and the number of partitions was explored via fault injections. Interestingly, a unified model that can quantify the effects of both SCU and MCU at an early design stage is not reported to date. One of the goals of our work aims at overcoming these limitations.

Early analysis of availability and reliability impacts of SEUs on a design can help designers developing more reliable and efficient systems while reducing the overall cost associated with their design. As spacecrafts are typically constrained by strict power budgets, too frequent scrubbing will drain energy. By contrast, less frequent scrubbing will allow accumulation of SEUs, which will eventually break TMR tolerance. Thus, analyzing trade-offs between the number of TMR partitions and scrub frequency is necessary. We propose a methodology based on Probabilistic Model Checking, to analyze this relationship at early design stages while analyzing reliability and availability of the design. Probabilistic model checking is a well known formal verification technique, mainly based on the construction and analysis of a probabilistic model, typically a Markov chain. The main advantage is that the analysis is exhaustive, and therefore results in numerically exact answers to the temporal logic queries, which contrasts discrete-event simulations \cite{PRISM:caseStudy}. Another advantage of this technique is its ability to express detailed temporal constraints on system executions in contrast to analytical methods. It is worth mentioning that formal verification is widely used in industry and government research agencies including Air Force Research Laboratory (AFRL),  NASA JPL, NASA Ames, NASA LaRC, National Institute of Aerospace (NIA) and so on,   for verification of safety-critical hardware and software systems \cite{formal1, formal2, formal3}.

In our previous work \cite{khaza,hoque2017formal}, we proposed a methodology that evaluates a Markov reward model (constructed from its high-level description) against performance-area-dependability tradeoffs using the PRISM model checker. In brief, we showed that there are cases where rescheduling of a Data Flow Graph (DFG) in conjunction with scrubbing can serve as a better fault-tolerant mechanism when compared with another fault-tolerant mechanism that combines the use of spare components with scrubbing. In addition, we also explored the relationship between fault detection coverage and scrubbing intervals. In this paper, we use a similar technique to extract the DFG from the high-level description of a system as explained in \cite{hoque2017formal} in detail. The main contribution of this paper is modeling of partitioned TMRs to explore the relationships between the scrub frequency and the number of TMR partitions in early design stages. Our modeling and analysis exploit the formal verification tool PRISM. This makes our work highly desirable since the use of formal verification techniques (especially model checking) in system design phases to verify the Design Assurance Level (DAL) compliance (as defined in the DO-254 standard \cite{miner2000case} for airborne electronic hardware) is highly recommended by NASA and Federal Aviation Administration (FAA) \cite{FAAFM}. Of course, performing the analysis only at early stage is not sufficient. Later in the design phases, hardware designers and radiation experts will need to perform additional tests like fault injection and beam testing in nuclear reactors. However, an early analysis in the design phase, such as our proposed approach, can reduce the overall design time, effort and cost. Hence we argue that our proposed technique \emph{bridges the gap} between FPGA designers, radiation experts and applied computer science using formal verification techniques. 

In our proposed approach, starting from a high-level description of a design (that employs partitioned TMR), the system is first modeled using the Continuous Time Markov Chain (CTMC) formalism. For CTMC modeling, we utilize the DFG of the system, the number of intended partitions, the scrub rate, and the failure rates that are obtained from a component characterization library. The model is then encoded using the PRISM modeling language. Properties related to the system's reliability and availability expressed using probabilistic temporal logic (in our case Continuous Stochastic Logic (CSL) \cite{Baier99approximatesymbolic}) are then verified automatically using the PRISM model checker \cite{PMC}, and the relationship between the scrub interval and the number of TMR partitions is assessed. In summary, our contributions in this paper are:

\begin{enumerate}

\item We propose formal models (using the CTMC formalism) of partitioned-TMR systems irrespective of the partition size (equal/non-equal sized). The proposed model captures both SCUs and MCUs. We limit our model up to SCUs and Double-Cell Upsets (DCUs).\\
\item We propose a methodology for early trade-off assessment of TMR partitioning and periodic blind scrubbing. The proposed modeling technique is \emph{modular}, which means each partition of the TMRed system is first modeled as a separate CTMC and then composed in parallel to a larger CTMC that models all the TMR partitions in the system. Since the approach is \emph{modular}, it is easily extendable to any number of partitions\footnote{However, large number of partitions may cause the well-known state-space explosion problem for model checking tools, we discuss this issue in detail in section \ref{ssec:scale}}.\\
\item Our CTMC formalization of the TMR partitioning is then encoded using the PRISM modeling language, and we utilize the Probabilistic Model Checking technique using the PRISM tool to perform quantitative assessments of an SCU prone, and an SCU and DCU prone FIR filter. Our analysis shows that increasing the number of TMR partitions also increases the design reliability and availability for the designs that are prone to both SCUs and DCUs (in the case when the failure of TMR voters are ignored). Moreover, the number of partitions also has a direct relationship with the configuration memory scrubbing frequency. Indeed such a relationship can be \emph{quantitatively} analyzed at early design stages based on the design requirements and constraints using the proposed methodology.\\
\item We also show that based on some given failure rate of voters (in contrast to the previous case where the failures of TMR voters were ignored), it is possible to identify the optimal number of partitions that offers the highest reliability. To our knowledge, this is the first unified model that captures SCUs, DCUs and voter failures for optimal TMR partitioning at early design stages.   
\end{enumerate}

The remainder of the paper is organized as follows. Section 2 reviews the related works in this area. Section 3 describes the background about SEU mitigation techniques and probabilistic model checking. The proposed methodology and modeling details are discussed in section 4, and in section 5, we present quantitative results obtained using our proposed methodology. Section 6 concludes the paper with proposed future research directions.
\begin{figure}[!t]
\centering
\includegraphics [height=10mm]{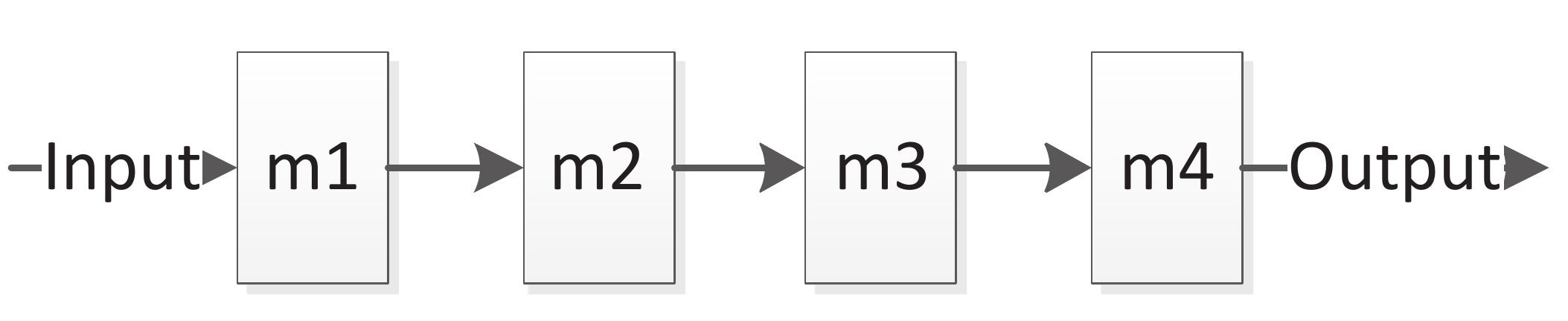}
\caption{Sample unmitigated circuit}
\label{fig:sample-ckt}
\end{figure}

\section{TMR partitioning and Related works}

\begin{figure}[!t]
\centering
\includegraphics [height=30mm]{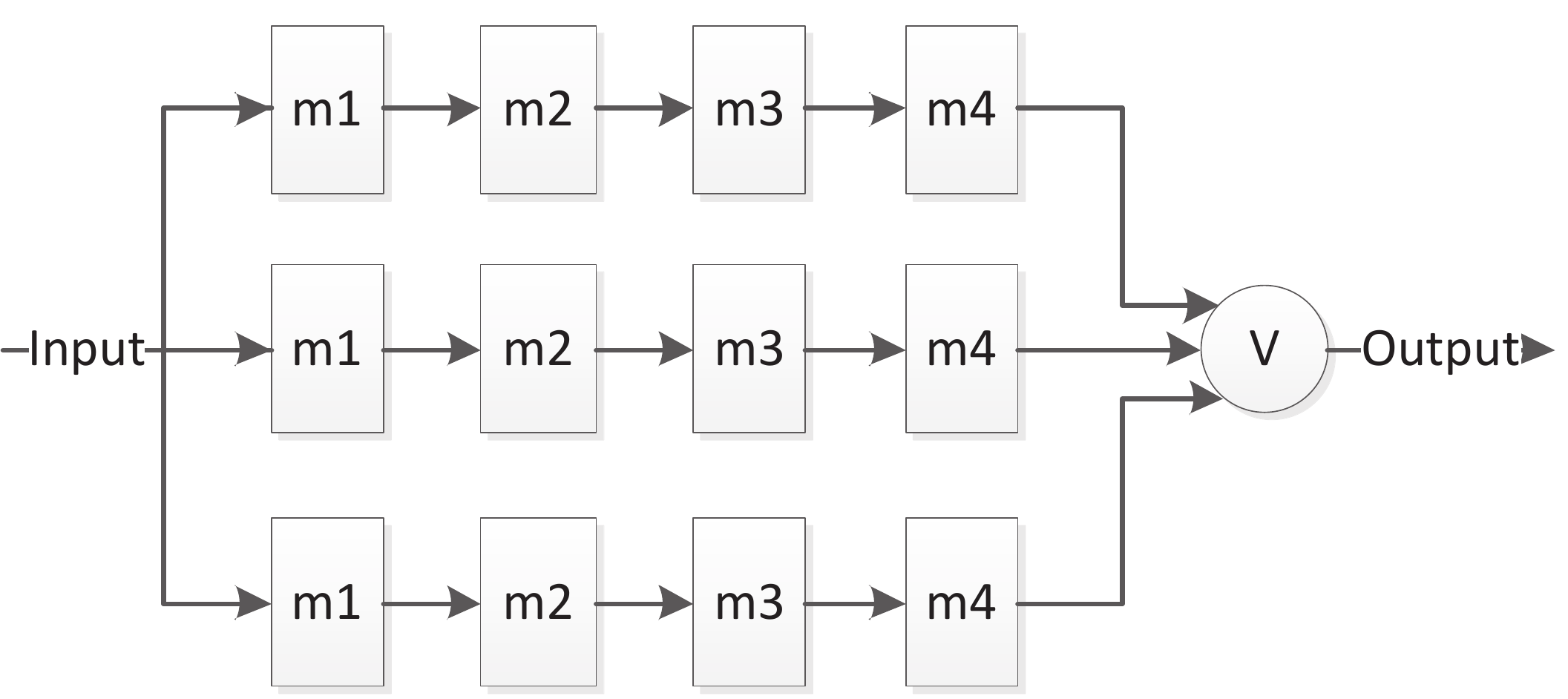}
\caption{TMRed version of the sample circuit}  \label{fig:TMR}
\end{figure}
\begin{figure}
\centering
\includegraphics [height=28mm]{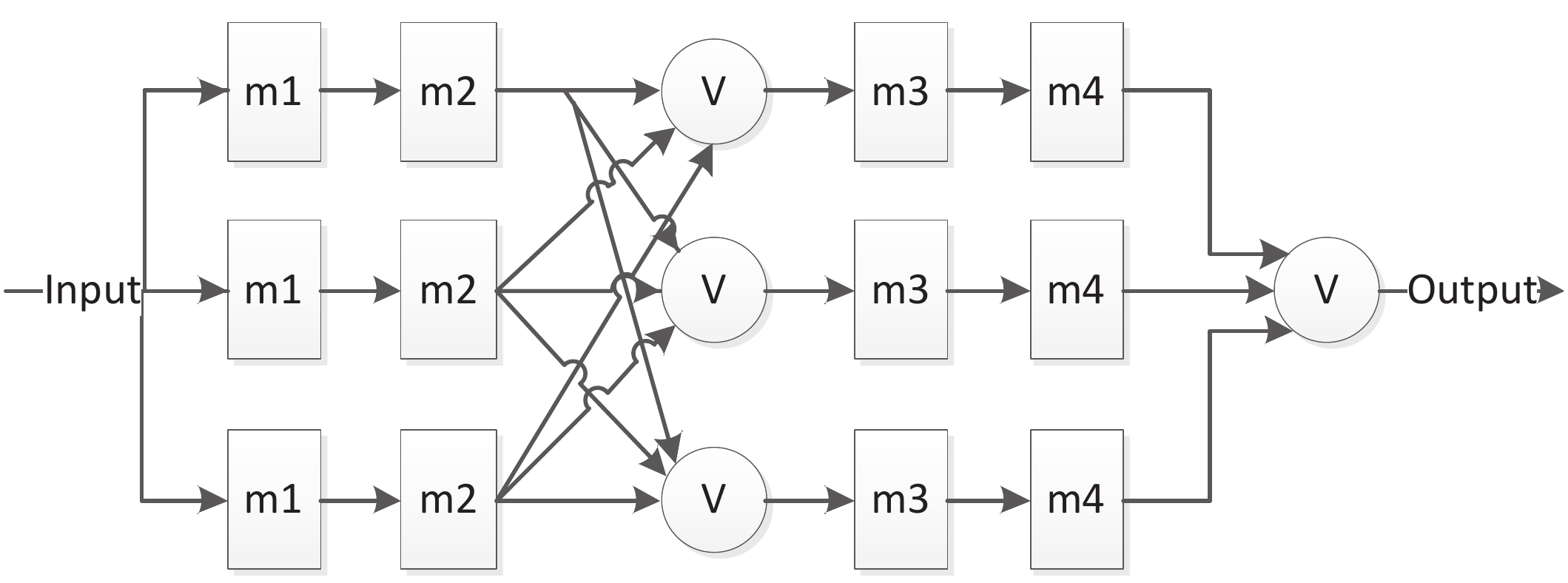}
\caption{The sample circuit with TMR divided into two partitions}  \label{fig:TMR_part}
\end{figure}

Traditional TMRed design can deal with a single fault at a time. Thus, faults in multiple redundant modules will break the TMR. For illustration, \figurename~\ref{fig:sample-ckt} shows a sample circuit (each box represents a module) and \figurename~\ref{fig:TMR} shows the TMR implementation of the sample circuit in the traditional way. While designing a system with traditional TMR, the components are triplicated and a majority voter is placed at the output of the circuit. The voter can provide correct outputs even if one of the branches (or domains) of the TMR is faulty. \figurename~\ref{fig:TMR_part} shows the same system implemented with partitioned TMR (as suggested in \cite{Related-2TMR}). In terms of dependability, each partition can be considered as a separate entity. This circuit will only fail if two or more domains in the same partition are affected by one or more faults. For example, upsets in module $m2$  in domain two (second row) and $m3$ in domain one (first row) will break a traditional TMR system, whereas it will get successfully masked in a system with partitioned TMR.

In FPGA configuration bitstream, a bit that is important for the functionality of a design is categorized as a \emph{critical~bit}. Using the component characterization library \cite{LibraryBasedSER}, the number of \emph{essential bits}, also known as \emph{potentially~critical~bits} to implement each component (adder, multiplier, etc.) in a target FPGA can be estimated. More accurate SEU susceptibility analysis can be performed using the fault injection techniques \cite{fault1,SEUFault}, however, for first-order worst-case estimation, it is valid to assume that all the \emph{essential bits} are considered as  \emph{critical bits}. In \cite{khaza, Khaza_MEMOCODE2014}, authors demonstrated how rescheduling of data-flow graphs can help to optimize the performability, and also showed the use of Erlang distribution for accurate modeling of scrubbing technique. In these works, to facilitate the early analysis, authors used the concept of \emph{characterization library} \cite{LibraryBasedSER}. In our proposed approach, we use the characterization library to calculate the failure rate of the TMR domains. It is worth mentioning that the proposed methodology is generic enough to be used with a different characterization library with more precise and accurate data (from radiation experts), without any major changes.

There are three main techniques to analyze SEU sensitivity in ASIC or FPGA based designs: 1) hardware testing with techniques such as particle beams and laser testing 2) fault injection emulation or simulation; and 3) analytical techniques. These three types of techniques are complementary, and they are typically applied at different design flow steps. The first two techniques are expensive in terms of cost and testing time. Moreover these two techniques often require a completed implementations \cite{ret1, ret5}. On the other hand, early analysis using analytical methods tends to be relatively less accurate in some aspects. Nonetheless, they can provide a much better controllability and observability, while enabling quick estimation of SEU susceptibility analysis, without the risk of damaging the devices \cite{sterpone2007experimental}. In addition, they can also capture features of the true test conditions that would be very hard to accurately reproduce when bombarding the circuit or while performing fault injection. Analytical estimation traditionally provides information at an earlier stage in the design cycle compared to the other two techniques.

Both academia and industry have heavily studied reliability and availability prediction of TMR. The effectiveness of different TMR schemes implemented with a different level of granularity has been evaluated experimentally (using beam test) and reported in \cite{rel-TMR-2015}. In \cite{Rel-TMR-AnalyticalModel-Zhong}, authors proposed an analytical model for systems with TMR, TMR with EDAC and TMR with scrubbing. They discuss the Markov modeling of these techniques throughout the paper, however, frequent voting or partitioning was not addressed. In \cite{AHS2013}, authors present a design flow to scrub each domain in a TMR independently to maximize the availability. In their approach, each partition is scrubbed on-demand when required. Since TMR is very expensive in terms of area and power, in \cite{rel-TMR-FineGrain} authors show TMR can be implemented only on selected portions of a design to reduce cost. Even though some level of reliability is sacrificed in this approach (compared to the approach where the whole system can be triplicated), in terms of area constraint, this tool may maximize the reliability.  The work in \cite{Related-2TMR} proposes a reliability model for partitioned TMR systems, but only for designs with equal sized partitions. Also, the possible failure of voters was not considered in any of the works mentioned above. 

A probabilistic model checking based approach for evaluating redundancy-based software system was proposed in \cite{remAfter}. The effect of domain crossing event and how to insert the voter cleverly in a hardware design was demonstrated in \cite{Related-3TMRoptimal}. In this work, the authors analyzed different partitioning schemes for the same design, and using the fault injection technique finds the optimal number of TMR partitions suitable for that design. Our work contrast all of these related works mentioned above. The proposed models can handle the effect of both SCUs and DCUs on designs, irrespective of the partition's size. In addition, the proposed model also considers the voter failures based on which optimal partitioning can be obtained.  
\section{Preliminaries}

\subsection{Probabilistic Model Checking}
Model checking~\cite{Clarke86automaticverification} is a well-established formal verification technique used to verify the correctness of finite-state systems. Given a formal model of the system to be verified in terms of labelled state transitions and the properties to be verified in terms of temporal logic, the model checking algorithm exhaustively and automatically explores all the possible states in a system to verify if the property is satisfiable or not. \emph{Probabilistic model checking} deals with systems that exhibit stochastic behaviour and is based on the construction and analysis of a probabilistic model of the system. We make use of CTMCs{, having both transition and state labels, } to perform stochastic modelling.
	
\begin{theorem}
{The tuple $C =(S, s_0, \Act, L, \Rt)$ defines a CTMC which is composed of a set of states $S$, an initial state $s_0 \in S$, a finite set of transition labels $\Act$, a labelling function $L: S\rightarrow 2^{AP}$ which assigns to each state $s \in S$ a subset of the set of atomic propositions $AP$ which are valid in $s$ and the transition rate matrix $ \Rt: S \times  S \rightarrow \Rt_{\geq 0}$. }The rate $\Rt(s,s')$ defines the delay before which a transition between states $s$ and $s'$ takes place. If $\Rt(s,s') \ne 0$ then the probability that a transition between the states $s$ and $s'$ is defined as $1 - e^{-{\footnotesize{\Rt}}(s,s') t} $ where $t$ is time. No transitions will  trigger if $\Rt(s,s') = 0$.
\end{theorem}

\begin{rmk} \label{sloop}
We allow self-loops in our CTMC model, and according to Definition 1, self-loops at state $s$ are possible and are modeled by having $\textbf{R}(s,s) > 0$. The inclusion of self-loops neither alters the transient nor the steady-state behavior of the CTMC, but allows the usual interpretation of Linear-Time Temporal (LTL) operators (we refer the interested reader to \cite{pnueli1977temporal} for more details about the syntax and semantics of LTL) like the \emph{next step} ($\mathcal{X}$) that we will exploit in Section~5 to check the correctness of the model.
\end{rmk}

In the Probabilistic model checking approach using CTMCs, properties are usually expressed in some form of extended temporal logic such as Continuous Stochastic Logic (CSL), a stochastic variant of the well-known Computational Tree Logic (CTL) \cite{Clarke86automaticverification}.  \\

\noindent  A CSL formula $\Phi$ defined over a CTMC $\mathcal{M}$ is one of the form:
\begin{center}

$\Phi ~::=~ true ~|~ a ~|~ \Phi \land \Phi ~|~ \neg \Phi ~|~ \mathcal{S}_{\bowtie p}(\Phi) ~|~ \mathcal{P}_{\bowtie p}(\phi)$\\
$\phi ~::=~ \mathcal{X} \Phi ~|~ \Phi \mathcal{U} \Phi ~|~ \Phi \mathcal{U}^{\leq t} \Phi$ \\
\end{center}

\noindent \emph{where $a \in AP$ is an atomic propositions, $p \in [0,1] $, $t \in \mathbb{R}_{> 0}$ and $ \bowtie \in \{<, \leq, \geq, > \}$. Each $\Phi$ is known as a state formula and each $\phi$ is known as a path formula.}\\

The detailed syntax and semantics of CSL can be found in \cite{Baier99approximatesymbolic}. In CSL, $\mathcal{S}_{\bowtie ~p} (\Phi)$ asserts that the steady-state probability for a $\Phi$ state meets the boundary condition $\bowtie~p$. On the other hand, $\mathcal{P}_{\bowtie~p} (\phi)$ asserts that the probability measure of the paths satisfying $\phi$ meets the bound given by $\bowtie ~p$. The meaning of the temporal operator $\mathcal{U}$ and $\mathcal{X}$ is standard (same as in LTL). The temporal operator $\mathcal{U}^{\leq t}$ is the real-time invariant of $\mathcal{U}$. Temporal operators like \emph{always} ($\Box$), \emph{eventually} ($\Diamond$) and their real-time variants ($\Box^{\leq t}$ and $\Diamond^{\leq t}$) can also be derived from the CSL semantics.  Below, we show some illustrative examples with their natural language translations:\\



\noindent 1. $ \mathcal{P}_{\geq 0.99} [\Diamond ~complete]$ - ``The probability of the system eventually completing its execution successfully is at least 0.99". \\
\noindent 2. $\mathcal{S}_{\leq 10^{-9}} [Failure] $ - ``In the long run, the probability that a failure condition can occur is less than or equal to $10^{-9}$ ". \\


\noindent In the PRISM property specification language \texttt{P, S, G, F, X} and \texttt{U} operators are used to refer to the $\mathcal{P, S, \Box, \Diamond, X}$ and $\mathcal{U}$ operator. In addition, PRISM also supports the expression $\mathcal{P = ? [\phi]}$ and $\mathcal{S = ?} [\Phi]$ in order to compute  the actual probability of the formula  $\phi$ and $\Phi$ being satisfied. PRISM also allows the use of customized properties using the \texttt{filter} operator: $filter (op, prop, states)$, where \emph{op} represents the filter operator (such as forall, print, min, max, etc.), \emph{prop} represents the PRISM property and \emph{states} (optional) represents the set of states over which to apply the filter.

\begin{figure*}[!t]
\centering
\includegraphics [width = \textwidth]{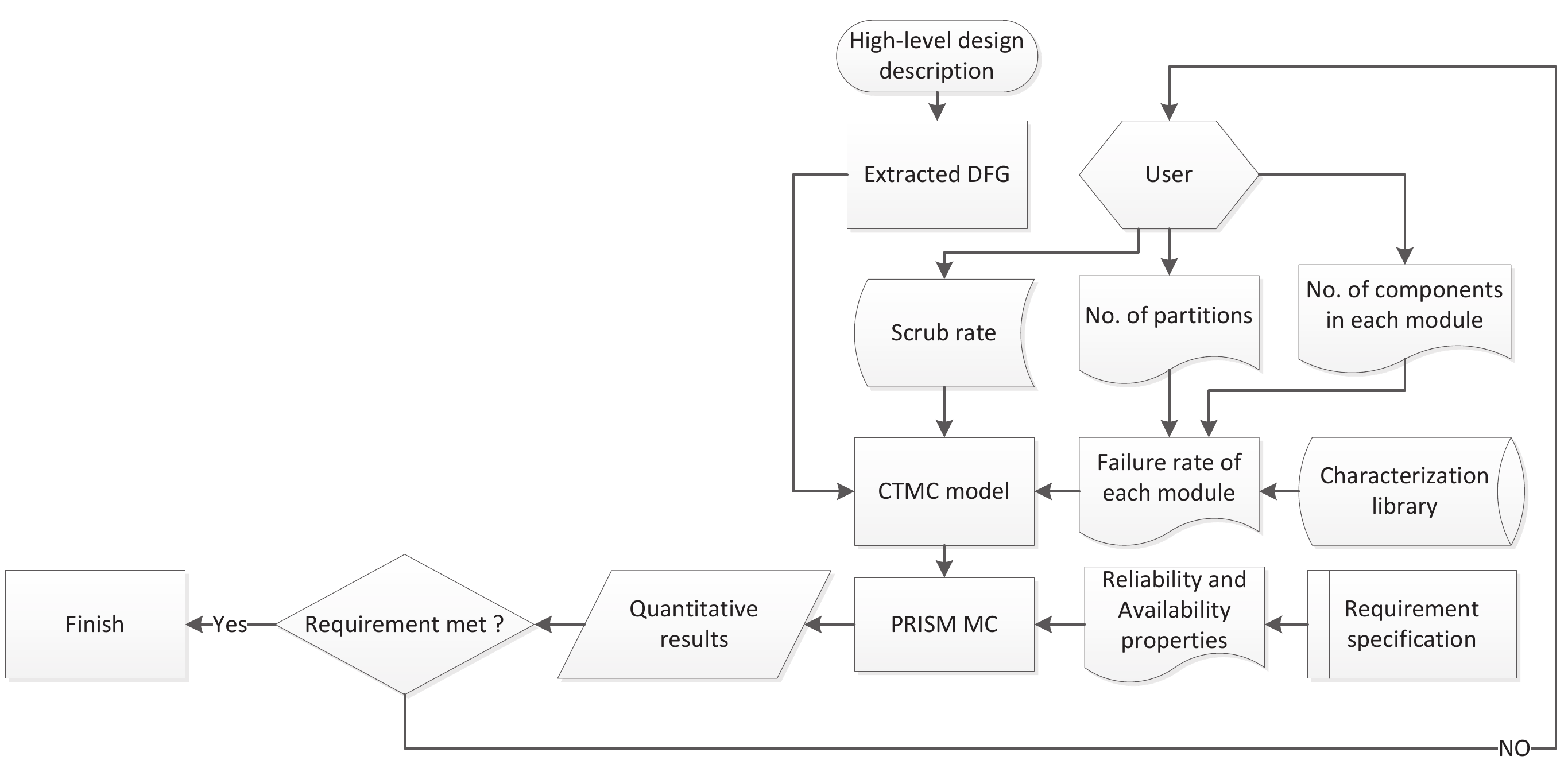}
\caption{Proposed methodology}  \label{fig:methodology}
\end{figure*}

\section{Proposed Methodology}
Figure \ref{fig:methodology} presents our proposed methodology. It starts from the high-level functional description of the system being designed, formulated in C/C++. The DFG is then extracted using the GAUT \cite{GAUT} tool. The DFG extraction part from a C/C++ code and the use of component characterization library is inspired from our previous work \cite{hoque2017formal}. Once the DFG is extracted, depending on the resource, performance or area constraint for hardware implementation, it can be scheduled using the appropriate scheduling algorithm. Since scheduling is out of the scope of this work, we assume a fully parallel implementation of the DFG for high performance. However, it is worth mentioning that the methodology will work irrespective of the scheduling approach.

Depending on the number and the size of partitions defined by the user, each domain in each partition can be represented as one or a collection of nodes (nodes in the graph represent a basic operation such as add, multiply, etc.). Each node can be implemented as a component in the FPGA. For clarity, we define the TMR module, domain and partitions in the context of our paper.\\
\begin{figure*} [!t]
	\centering
	\includegraphics[width = \textwidth]{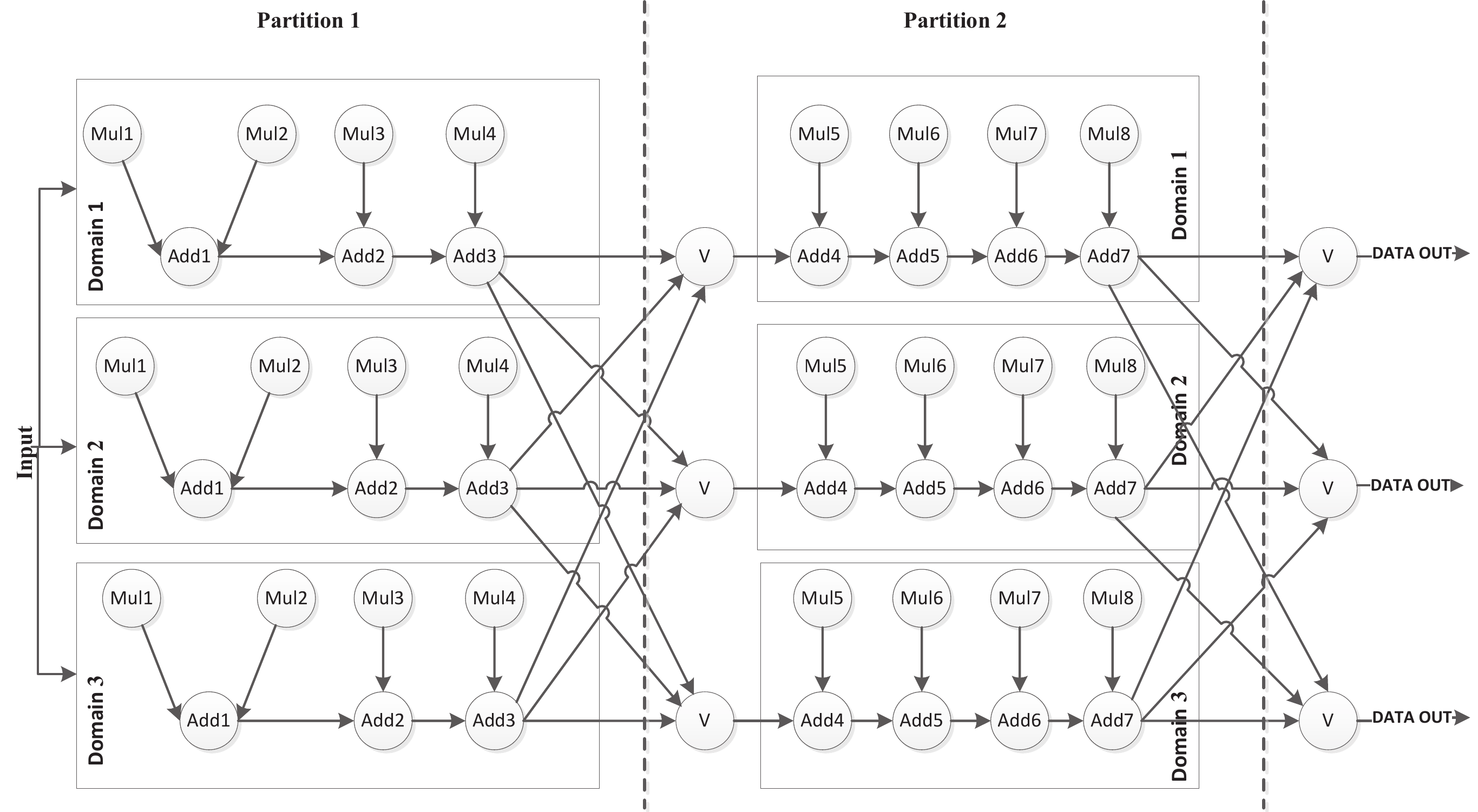}
	\caption{TMR implementation of the DFG of an 8-tap FIR filter with two partitions}\label{DFG_FIR}
\end{figure*}

\noindent \textbf{Module:} A module (or component) in a TMR refers to the basic operations as they appear in the C/C++ code of the design and the obtained DFG. The failure rate calculation of a module is based on the component characterization library from \cite{LibraryBasedSER}. As mentioned earlier, any other component characterization library (such as those obtained by fault injection or beam testing) can also be used with our methodology without any major changes.\\

\noindent \textbf{Domain:} A TMR domain (or branch) consists of one or more modules. An unpartitioned TMR will have three replicated domains, whereas a partitioned TMR will have three redundant domains in each partition. Based on this  fact, each domain in the same partition has an equal failure rate. In contrast, domains from different partitions may have equal or different failure rates depending on the size of partitions. Since the input of each domain is connected to the output of a voter from the previous partition (it is well known that if communication between systems is a single line it becomes a single point failure that can be avoided by triplicated wires carrying the 2 or more out of 3 code), hence the failure rate of a domain can be expressed as:

\begin{equation}
\lambda_{domain} = \sum^{k}_{i=1} \lambda_{i} + \lambda_{voter} \label{domain}
\end{equation}

\noindent where, $k$ is the total number of modules in a domain, $\lambda_{i}$ is the failure rate of module $i$, and $\lambda_{voter}$ is the failure rate of a voter. Note that, since there is no voter included in the first partition, hence $\lambda_{voter} = 0$ for the domains of first partition. The last three voters voting on the final output also can be modeled (with an extra partition) using Equation 1 by replacing $\sum^{k}_{i=1} \lambda_{i} = 0$. \\


\noindent \textbf{Partition:} A TMR partition refers to the logical partitioning of the TMRed design. Each partition may have equal or unequal size.  Voters are inserted after each partition to vote on the output from the domains. Since in a partition a domain is replicated three times, hence the upset rate of a (fully operational) partition is equal to $3*\lambda_{domain}$.\\

For illustration, in \figurename~\ref{DFG_FIR} the partitioning of a DFG representing an 8-tap FIR filter is shown. All the domains in partition-1 have four multipliers and three adders. On the other hand, each of the domains that are part of partition-2 contains four multipliers, four adders and three voters. The failure rate for a module is calculated using the following equation:

\begin{center}
$\lambda_{module} = \lambda_{bit} \times Number~of~critical~bits$
\end{center}

\noindent For our experiments, $\lambda_{bit} = 7.31 \times 10^{-12}$ SEUs/bit/sec where $\lambda_{bit}$ represents the SEU rate for the Higher Earth Orbit, and the \emph{number of critical bits} in a module is obtained from the characterization library. It is important to mention that an SEU can cause either an SCU or an DCU. Hence, $\lambda_{domain}$ needs to be adjusted accordingly. The simplest way is to multiply the $\lambda_{domain}$ with the SCU and DCU coefficient, $\alpha_{SCU}$ and $\alpha_{DCU}$ respectively.

Based on the calculated failure rate of each domain, the number of partitions and the user defined scrub rate, a CTMC model of the system is then built and encoded in PRISM modeling language. Different reliability and availability properties are then verified using the PRISM model checker to assess if the design meets its requirements based on the provided quantitative results. If  requirements are not met, the number of partitions or the scrub frequency is then modified, and the analysis is performed again.


\subsection{Formal Modeling of partitioned TMR}
We start with formalizing individual TMR partitions, and then show how to construct the whole partitioned system. A system with $N$ partitions can be defined by a set:

\begin{equation*}
	P = \{{P_1, P_2, .... , P_N }\}
\end{equation*}

Here each $P_i \in P$ represents a TMR partition. Each of these partitions are prone to SCUs. However, as mentioned earlier an SEU can flip two or more bits simultaneously in the FPGA configuration bitstream, inducing MCUs in TMR partitions. This situation is more common in a harsh radioactive environment such as in outer space. As mentioned earlier, in this paper we limit our modeling to Double-Cell Upsets (DCUs). Also, for the proposed models, we consider the following assumptions:\\

\noindent \emph{Assumption 1}:~For the SCU model, each domain in the TMR may fail independently. The time-to-failure due to a configuration bit flip(s) (either inducing an SCU or DCU) is exponentially distributed. The exponential distribution is commonly used to model the reliability of systems where the failure rate is constant. The \emph{scrub} interval is assumed to follow an exponential distribution as well, with a rate, $\mu$ = $1/\tau$, where $\tau$ represents the scrub interval. It is also possible to approximate the deterministic scrub interval using the Erlang distribution as shown in \cite{Khaza_MEMOCODE2014, hoque2015towards}.  \\

\noindent \emph{Assumption 2}: The design employs the blind scrubbing technique \cite{XilinxScrub, intvsExtScrub}. Blind scrubbing is a very popular and reliable scrubbing strategy that requires no additional detection algorithm before fixing the configuration memory upsets.\\

As according to the fault-free design structure, there is no electrical node in the logical netlist stemming from one domain that is re-converging in another domain, the domains are physically independent, except if a set of flipped configuration bits (one or more than one) introduce a short between two nets belonging to two distinct domains. This is a physical design issue with the FPGA and how it is placed and routed. We can either neglect this possibility as its occurrence probability is expected to be low, or do careful physical design such that the probability of the event becomes negligible. The MCU modeled in the paper considers the scenario where a single SEU may cause a double-bit upset causing two domains to fail simultaneously. However, for SCUs this is not generally the case (subject to the above discussion). Hence, according to assumption 1, domains should fail independently (since they are structurally independent). Since failure of a single or multiple module (component) in the same domain can lead to the domain failure, hence the failure rate of a domain is equivalent to the summation of the number of essential bits from all the modules in that domain (as shown in Equation 1).

\subsubsection{Modeling of Single-Cell Upsets (SCUs)}

\noindent Each TMR partition $P_i \in P$ which is prone to SCUs can be described as a CTMC.

\begin{theorem}[SCU-prone partition model]
An SCU prone $i^{th}$ TMR partition is defined by a CTMC $\mathcal{M}_{i}^{scu} = (S_{scu}, s_0, \Act_{scu}, L_{scu}, \Rt_{scu})$, where $S_{scu} = \{3,2,1\}$, $ s_0 = \{3\}$, $\Act_{scu} = \{scu_1, scu_2, perform\_scrub\}$, $L(3) = \{operational\}$, $L(2) = \{degraded\}$, $L (1) = \{failed\}$ and \[
\mathbb{R}_{scu}
=\begin{bmatrix}
    \mathbb{R}(1,1) & \mathbb{R}(1,2) & \mathbb{R}(1,3) \\
    \mathbb{R}(2,1) & \mathbb{R}(2,2) & \mathbb{R}(2,3)   \\
    \mathbb{R}(3,1) & \mathbb{R}(3,2) & \mathbb{R}(3,3)
\end{bmatrix}
=\begin{bmatrix}
    0 & 0 & \mu \\
    2\lambda & 0 & \mu   \\

    0 & 3\lambda & \mu
\end{bmatrix}
\]
\end{theorem}

\begin{figure}[!t]
\centering
\includegraphics [width = 0.6\textwidth] {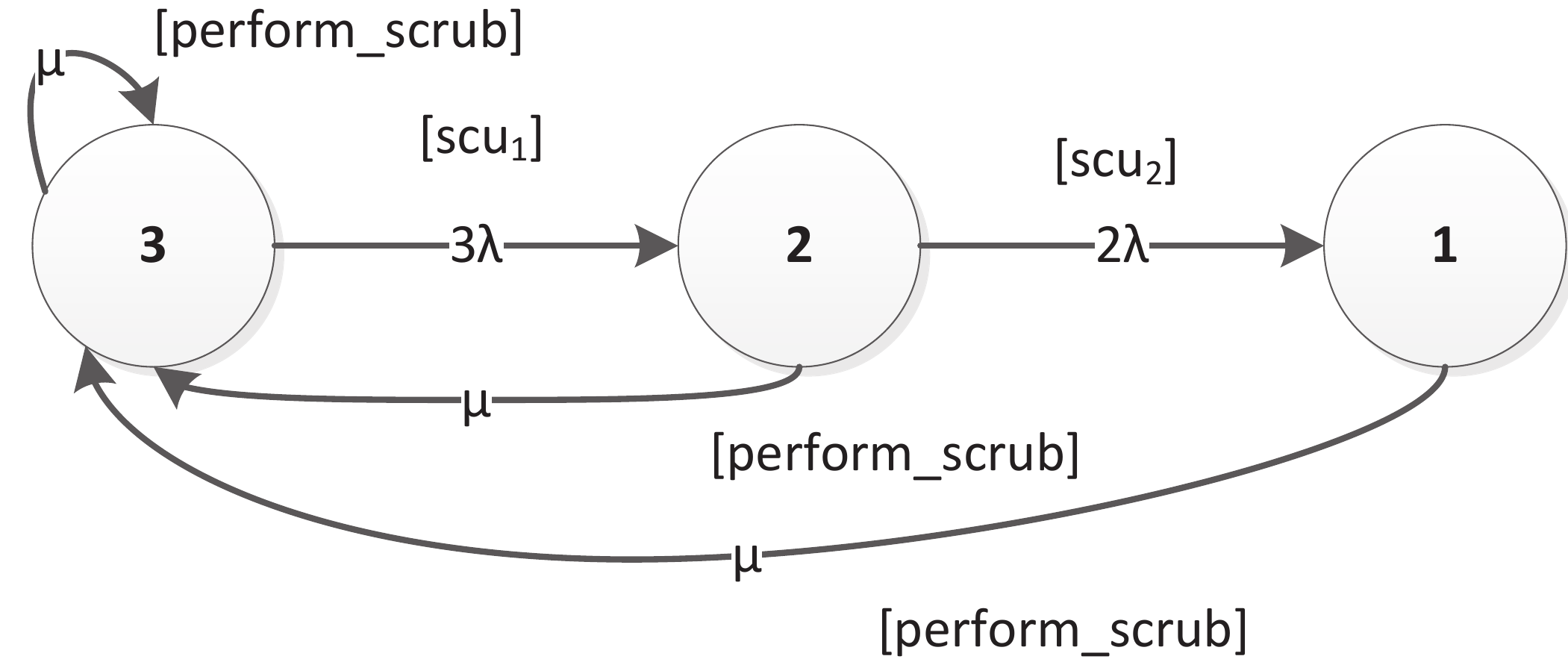}
\caption{Markov model of TMR with repair}  \label{fig:TMR_Markov}
\end{figure}

\figurename~\ref{fig:TMR_Markov} shows an SCU prone TMR partition model with scrubbing. We removed the state labels from the figure and the transition labels are specified inside square brackets ([~]) for clarity. Each node in the model denotes the current state of a domain in that \emph{specific} partition: state 3 ($operational$) represents the state in which all domains are operating correctly (all the modules are fault-free), state 2 ($degraded$) represents a state where one out of three domains is operating incorrectly (in one of the domains at least one of the modules is faulty), but the output is still not erroneous, and state 1 ($failed$) represents a \emph{failure state} in which two or more domains are operating incorrectly (more than one module is faulty in two or more domains). Since it is a TMR system, 2 out of 3 domains need to be working at a time.

In this model, $\mu$ represents the scrub rate. Note that, since we consider periodic blind scrub for this paper, the whole system gets scrubbed periodically. This is reflected in the model by the $perform\_scurb$ transition with the scrub rate $\mu$ (irrespective of its current state).

From state 3, the system can move to state 2 if it encounters an SEU with the $scu_1$ transition and the rate of this transition is $3*\lambda_{domain}$.  From state 2, the system has two options: (1) the system can get scrubbed and go back to state 3; or,  (2) another module in a fault-free domain of the TMR can fail --- which will lead the system to state 1 with a transition label $scu_2$ and transition rate $2*\lambda$. Once the system enters the state 1, which is a failed state, it will remain in that state until the system gets scrubbed eventually and comes back to state 3 with the scrub rate $\mu$. Since in our modeling, each TMR partition is modeled as a separate CTMC that is equivalent to the model shown in \figurename~\ref{fig:TMR_Markov}, this allows us to model systems with either equal or unequal sized partitions based on the fact that failure rate of a partition is reflected in its CTMC model. Once all the partitions are modeled as separate CTMCs, we construct the model of the overall \emph{partitioned} scrubbed-TMR system from the parallel composition~\cite{concurrency} of those CTMCs.

\begin{equation*}
  \mathcal{M}_{system} = \mathcal{M}_{1} \parallel \mathcal{M}_{2} \parallel .... ~{\mathcal{M}_{N} }
\end{equation*}

\begin{theorem}[Parallel composition] \label{def:synch}

Given two CTMCs $\mathcal{M}_1 = (S_1, s_{01}, \Act_{1}, L_1,\Rt_1)$ and $\mathcal{M}_2 =(S_2, s_{02}, \Act_{2}, L_2,\Rt_2)$, the parallel composition of $\mathcal{M}_1$ and $\mathcal{M}_2$ is the CTMC $\mathcal{M} = \mathcal{M}_1 || \mathcal{M}_2 = (S_1 \times  S_2, (s_{01}, s_{02}), \Act_1 \cup \Act_2, L_1 \cup L_2,\Rt)$ where,
\begin{equation*}
\Rt=
\begin{cases}

		
 \frac{s_1 \xrightarrow{\alpha, \lambda_1} s_1' ~ \text{and}~ s_2 \xrightarrow{\alpha, \lambda_2} s_2' }{(s_1,s_2) \xrightarrow{\alpha,\lambda_1\times \lambda_2} (s_1',s_2')}, & \mbox{if } \Act_{1} = \Act_{2} = \alpha\text{(Full synchronization)}, \\ \\
  \frac{s_1 \xrightarrow{\alpha_1, \lambda_1} s_1'}{(s_1,s_2) \xrightarrow{\alpha_1,\lambda_1} (s_1',s_2)},
			\frac{s_2 \xrightarrow{\alpha_2, \lambda_2} s_2'}{(s_1,s_2) \xrightarrow{\alpha_2,\lambda_2} (s_1,s_2')}, & \mbox{otherwise} \text{(Interleaving synchronization)}.
\end{cases}
\end{equation*}
and $s_1,s_1' \in S_1$, $\alpha_1 \in \Act_1$, $\alpha_2 \in \Act_2$, $\alpha \in \Act_1 \cap \Act_2$, $\Rt_1(s_1,s_1') = \lambda_1$, $s_2,s_2' \in S_2$, $\Rt_2(s_2,s_2') = \lambda_2$.

\end{theorem}

\begin{figure}[!t]
\centering
\includegraphics [width=0.7\textwidth]{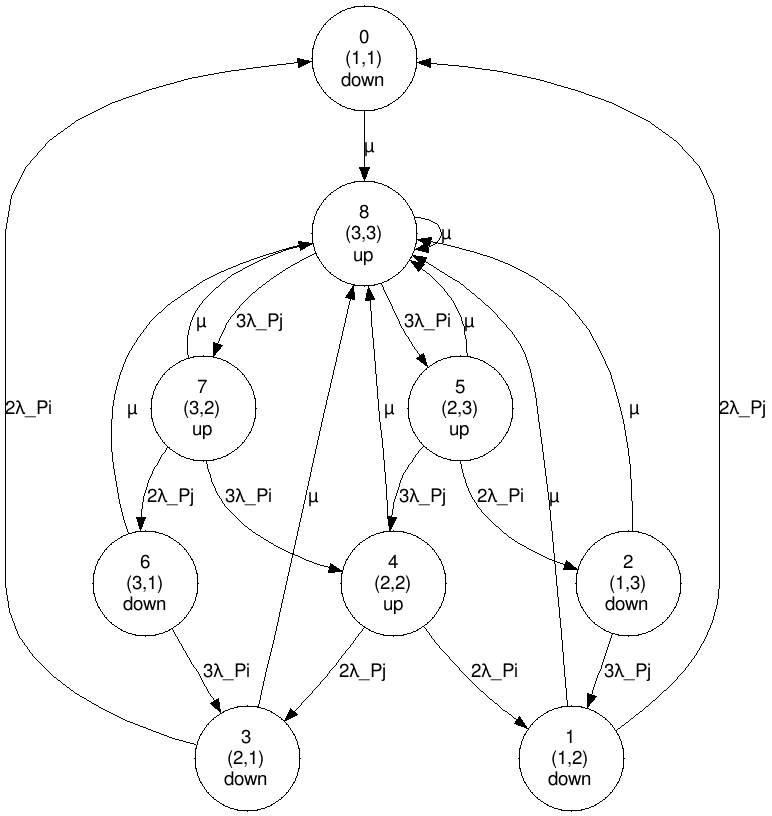}
\caption{SCU Markov model of TMR with two partitions}  \label{fig:Markov_two_partition}
\end{figure}


\figurename~\ref{fig:Markov_two_partition} shows the resulting composed CTMC for a TMR-scrubbed system with two partitions where $\lambda_{Pi}$ and $\lambda_{Pj}$ represent the failure rate of left and right partitions. Transition labels have been omitted from the figure for clarity. We further map the states as either $up$ or $down$ using the following function:

\begin{equation*}
	up = \begin{cases}
	1, & \mbox{if~} L(s)_{p_{i}} \wedge L(s)_{p_{i+1}} \wedge ..... \wedge L(s)_{p_{N}}  = \mathit{operational} \vee \mathit{degraded},\\
	0, & \mbox{otherwise}.\\
	\end{cases}
\end{equation*}

which means that if the current state of any of the partitions are labelled as $operational$ or $degraded$, then we classify those corresponding states as $up$, otherwise $down$. Hence, in \figurename~\ref{fig:Markov_two_partition}, $up = \{4,5,7,8\}$ and $down = \{0,1,2,3,6\}$


\subsection{Modeling of Multiple-Cell Upsets (MCUs)}

An SEU that causes a DCU invokes failures in multiple TMR domains simultaneously and each of those domain may belong to the same, or to the separate partitions. Hence, we need to consider two cases for modeling them: (1) DCUs in the same partition and (2) DCUs in separate partitions. To achieve the first goal, we enhance the model presented in \figurename~\ref{fig:TMR_Markov} and define them as follows:
\begin{figure}[!t]
	\centering
	\includegraphics [width = 0.6\textwidth] {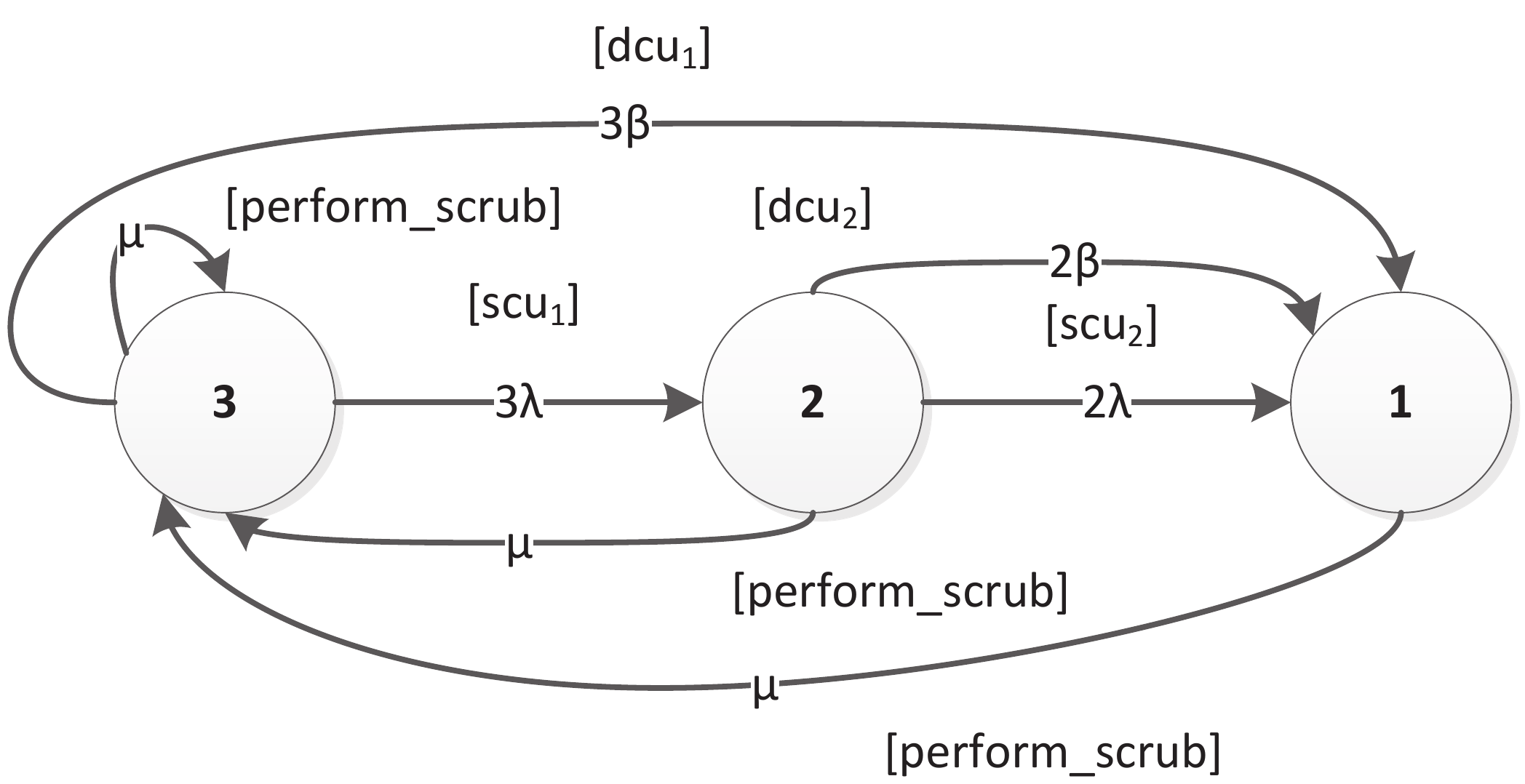}
	\caption{DCU model of TMR with repair}  \label{fig:TMR_DBU}
\end{figure}
\begin{theorem} [SCU and DCU prone combined model] \label{def:SCUDCU}
An $i^{th}$ TMR partition that is prone to both SCUs and DCUs is defined by a CTMC $\mathcal{M}_{i}^{comb} = (S_{comb}, s_0, \Act_{comb}, L_{comb}, \Rt_{comb})$, where $S_{comb} = \{3,2,1\}$, $ s_0 = \{3\}$, $\Act_{comb} = \{scu_1, scu_2, dcu_1, dcu_2, perform\_scrub\}$, $L(3) = \{operational\}$, $L(2) = \{degraded\}$, $L (1) = \{failed\}$ and \[
	\mathbb{R}_{comb}
	=\begin{bmatrix}
	\mathbb{R}(1,1) & \mathbb{R}(1,2) & \mathbb{R}(1,3) \\
	\mathbb{R}(2,1) & \mathbb{R}(2,2) & \mathbb{R}(2,3)   \\
	\mathbb{R}(3,1) & \mathbb{R}(3,2) & \mathbb{R}(3,3)
	\end{bmatrix}
	=\begin{bmatrix}
	0 & 0 & \mu \\
	2\lambda + 2\beta & 0 & \mu   \\
	3\beta & 3\lambda & \mu
	\end{bmatrix}
	\]
\end{theorem}

We refer to this model as the ``$combined~model$", since it captures both SCUs and DCUs (in the same partition) .

Let us first consider the case where a DCU causes failure of two domains, but in the same partition. The parameter $\beta$ represents the DCU rate of a domain pair, which corresponds to a situation causing a domain failure while also causing failure of another domain in the same partition (that would happen if an ionizing particle first hit a domain and then induces an ionized track spreading to a second domain to also make it fail). For instance, let us consider the DCU case where domain 1 fails first in a partition and then causes a failure to domain 2 of the same partition with a rate $\gamma$ (as paticle upsets take place on a very short time scale, the two domains are assumed to fail at the same clock cycle). The reverse could also happen, which means domain 2 can fail while invoking a failure to domain 1 with a rate $\gamma$. So, for the pair domain 1 and 2, the rate at which either of them fails in the same cycle as to the other one is $\gamma+\gamma$. As defined, the two considered events are disjoint and their rates can be summed accurately. In our model, we combined and express them together, which means $\beta = \gamma+\gamma$. Similarly, we have two more domain pairs to consider in a TMR partition (irrespective of their order since we combine the rates), which are domain 1 and domain 3, and domain 2 and 3. In that context, \figurename~\ref{fig:TMR_DBU} shows the Markov model of a TMR partition that can handle both SCUs and DCUs. Compared to the SCU partition model, the main changes are two new transitions, from state 3 to state 1 ($[dcu_1]$) and from state 2 to state 1 ($[dcu_2]$). The $[dcu_1]$ transition with the rate $3 \beta$ represents the phenomenon that at least one of the DCUs in one of the three domain pairs (in the same partition) will cause a TMR failure. Similarly, the $[dcu_2]$ transition also represents another domain pair failure in the same partition.

The second case models a DCU affecting two separate partitions (in addition to the DCUs affecting in a same partition). This case is significantly more complex, and to model this we need to introduce new transitions (in addition to those defined in Definition \ref{def:SCUDCU}) in separate partitions . It requires the use of \emph{pairwise} synchronization (via \emph{parallel composition} as defined in Definition~\ref{def:synch}) of associated transitions in different partitions to represent a simultaneous failure due to a DCU in different partitions.




\begin{figure}[!t]
\centering
\includegraphics [width = \textwidth] {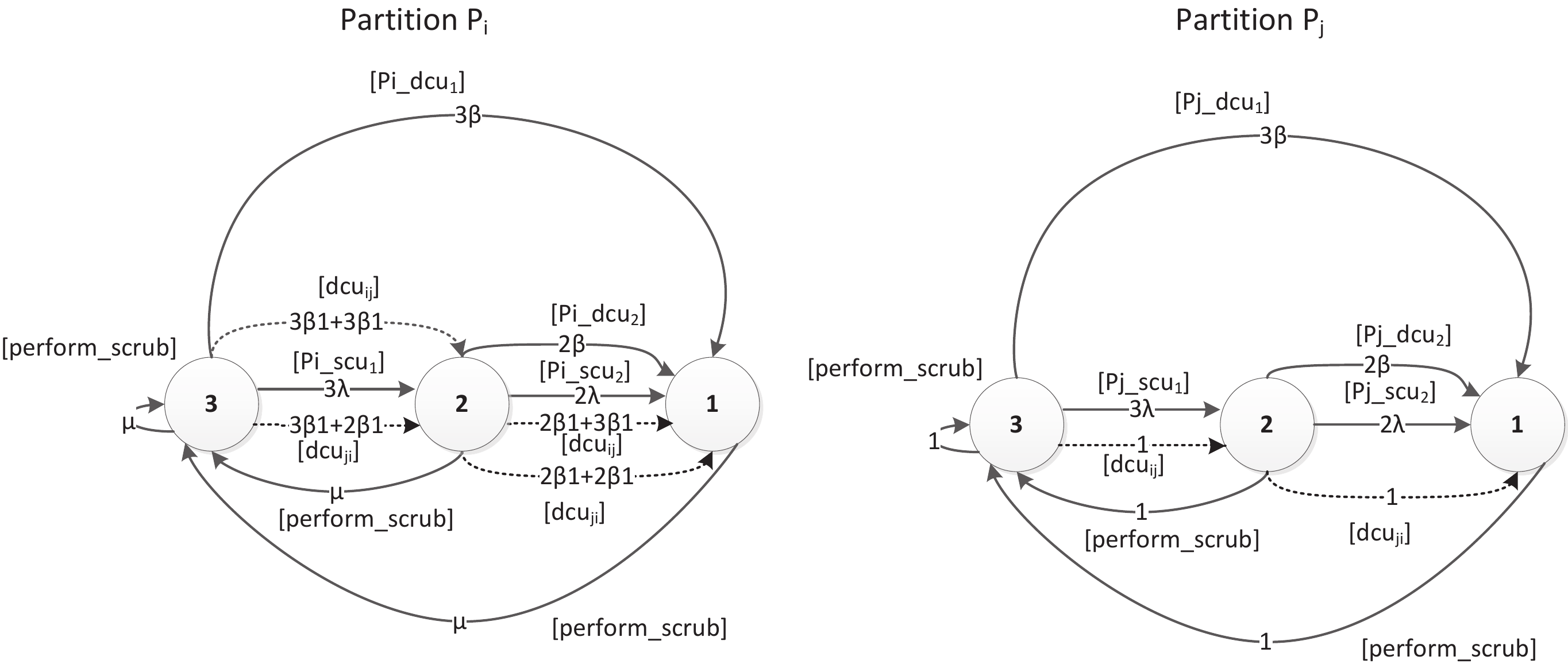}
\caption{Model for synchronization of DCUs in separate domain (two partitions)}  \label{fig:TMR_DBUSynch}
\end{figure}

Figure~\ref{fig:TMR_DBUSynch} shows how we extend the ``$combined~model$" (that was shown in Figure~\ref{fig:TMR_DBU}) for incorporating DCUs in distributed domains (shown by dotted arrows) for the case where a design has two partitions (corresponds to the layout shown in Figure \ref{DFG_FIR}). In this extended model, we represent DCUs in the same partition using the previously defined parameter $\beta$ and DCUs in separate partitions using another parameter $\beta1$. To define $\beta1$, we consider that domain 1 in $P_i$ can fail while invoking another failure of domain 1, domain 2 or domain 3 in partition $P_j$ with a rate $\gamma1+\gamma1+\gamma1$. In our model, this is represented as $\beta1=\gamma1+\gamma1+\gamma1$.  So the rate at which the three domains in partition $P_i$ will fail due to a DCU while causing a failure to a domain in partition $P_j$ is $3\beta_1$. Similarly, the rate that either of the three domains in partition $P_j$ will fail due to a DCU while invoking a failure to a domain in partition $P_i$ is also $3\beta_1$. In Figure~\ref{fig:TMR_DBUSynch}, the transition $[Pi\_dcu_1]$ and $[Pj\_dcu_1]$ represent a DCU in partition $P_i$ and $P_j$ respectively (DCUs in the same partition). In contrast, the $[dcu_{ij}]$ and $[dcu_{ji}]$ transitions (triggers based on the current state) in partition $P_i$ is synchronized using the same transition label ($[dcu_{ij}]$ and $[dcu_{ji}]$) in partition $P_j$ with the rate $1$ (refer to Remark~\ref{rmk1}), which depicts DCUs in two separate partitions. Let us consider a case where both partitions have three operational domains, hence both Markov chains are in state 3. For a DCU in separate partitions, either encountered by a domain in partition $P_i$ or $P_j$, both Markov chains will move to state 2 simultaneously (this is synchronized using the same label $[dcu_{ij}]$ in both partitions). If the partition $P_j$ is in state 2, and partition $P_i$ is in state 3, then the left Markov chain will move to state 2 and the right Markov chain will move to state 1 simultaneously, and this has been synchronized using the same label $[dcu_{ji}]$.

\begin{rmk} \label{rmk1}
Please note that, for full synchronization, as in Definitions~\ref{def:synch}, the rate of a synchronous transition is defined as the product of the rates for each transition. We need to synchronize the scrub transition ``perform\_scrub" between the TMR partition models. For example, the intended scrub rate ($\mu$) is specified in full for the scrub transitions in one of the partitions (in the first partition as shown in Figure~\ref{fig:TMR_DBUSynch}), and the rate of other scrub transition(s) in rest of the partition models (in second partition as shown in Figure~\ref{fig:TMR_DBUSynch}) are specified as 1. Similarly we synchronize transitions for modeling DCUs in separate domains.
\end{rmk}

\begin{figure*}[!t]
	\centering
    \includegraphics [width=\textwidth]{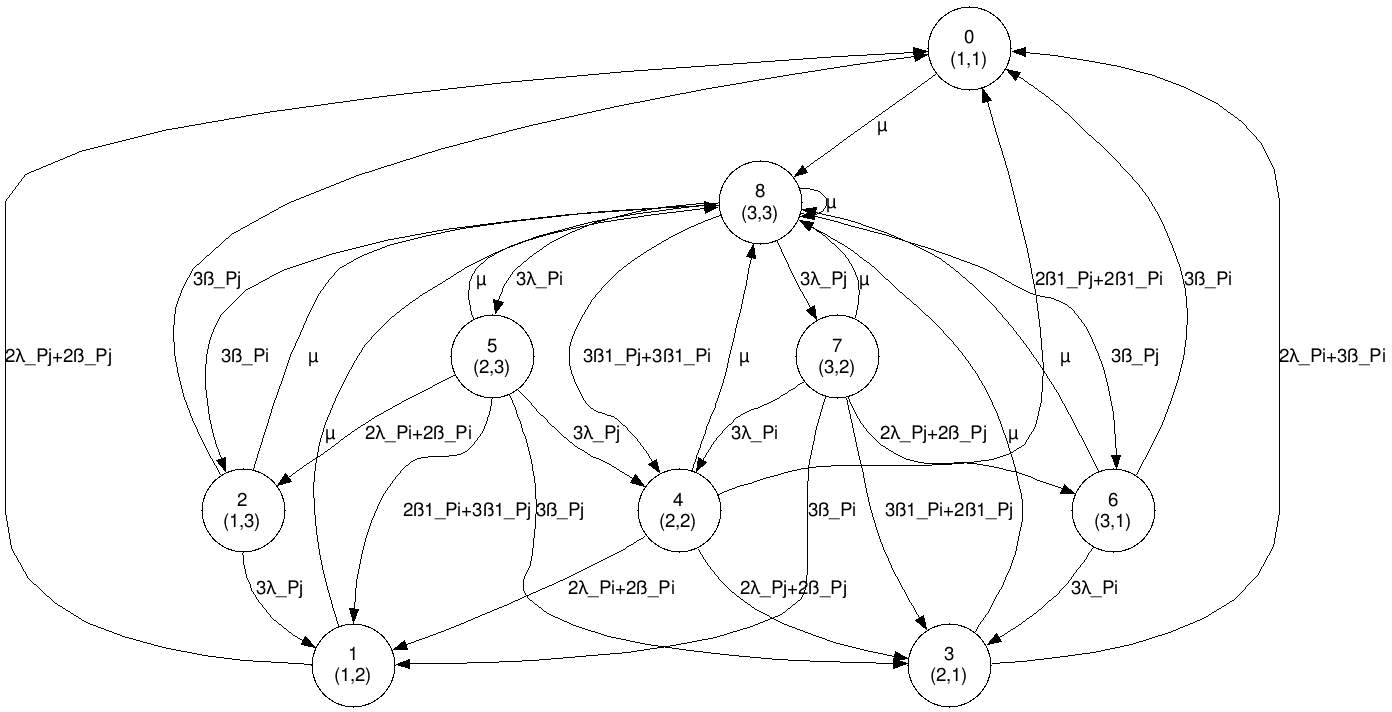}
	\caption{Combined Markov model of TMR with two partitions}  \label{fig:Markov_DBU}
\end{figure*}

\figurename~\ref{fig:Markov_DBU} shows the \emph{combined mode}l after parallel composition of the partitions (refer to Definition~\ref{def:synch}) that encapsulates the effect of both: SCUs and DCUs (same and separate domains). In this model, $\beta_{Pi}$ and $\beta_{Pj}$ represent the DCU rate (DCUs in the same partition) of a domain in the first and second partition respectively. The parameter $\beta1_{Pi}$ and $\beta1_{Pj}$ represent DCUs in respective separate partitions. 

For example, in state 8 both partitions are working fine. However, if one of the domains in the partitions encounters an SCU, then the system can move to either state 5 or state 7, depending on the location of the domain. Also, if the system is in state 8, and a DCU occurs in any domain of either partitions (DCU in separate partitions), it will trigger another domain failure in the other partition simultaneously. This leads to a path from state 8 to state 4 with the rate $3*\beta1_{Pj}+3*\beta1_{Pi}$. The rest of the two transitions from state 8 to state 2 and state 6 represent DCUs in the same partition with associated rates.

For our analysis, we developed Markov models for four design options, starting from no partition\footnote{no partition (unpartitioned TMR) actually refers to 1 partition} up to eight partitions. The complexity of these models in terms of total number of states and total number of transitions is shown in Table \ref{tab:model_stat}. As observed from Table \ref{tab:model_stat}, with increasing number of partitions, modeling of TMR gets very complicated since increasing number of transitions are required to model synchronized  DCUs in separate domains. We were able to keep such modeling manageable since we define each partitions separately as \texttt{modules} in the PRISM language and utilize the PRISM model checking tool for parallel composition of \texttt{modules} (representing TMR partitions) for generating the complete model for analysis. Please note that \texttt{modules} in PRISM language and modules in a TMR should not be confused.  We refer to \cite{PMC2prism} for details about the PRISM modeling language. Similarly, $N$ partitions can be modeled using our methodology by adding new \texttt{modules} to the PRISM code.

\begin{table}
\centering
\caption{Model construction statistics}  \label{tab:model_stat}
    \begin{tabular}{|c|c|c|c|}
    \hline
  \textbf{No. of } & \textbf{No. of } & \textbf{No. of transitions} & \textbf{No. of transitions}\\
  \textbf{partitions} & \textbf{states}& \textbf{(SCU only model)} & \textbf{(combined model)} \\ \hline
  1 & 3 & 5 & 11\\ \hline
  2 & 9 & 26 & 36 \\ \hline
  4 & 81 & 362 & 578 \\ \hline
  8 & 6561 & 47858 & 129506 \\ \hline
  \end{tabular}
\end{table}

%
%
%
%
%

\section{Quantitative Analysis of an FIR Filter}

Filters are commonly used in digital communication systems for different purposes, such as for equalization, signal separation, noise reduction and so on. Communication is a fundamental issue for all space-borne applications ranging from satellites to unmanned missions. That is why digital filters have an important role to play for such systems \cite{FIRdsp}. To illustrate the applicability of our approach, we analyze an 8-bit 64-tap FIR filter (the target platform is a Xilinx Virtex-5 SRAM-based FPGA) using both, the SCU model and the combined model for a different number of partitions. FIR filters \cite{FIR-3} are widely used in space applications for their excellent stability and simplicity of their implementation according to the given response. An N-tap discrete finite impulse response (FIR) filter can be expressed as follows:

\begin{equation*}
  y[n]=\sum_{i=0}^{N-1}x[n-i]\cdot h[i]
\end{equation*}

\noindent $x[~]$, $y[~]$ and $h[~]$ are the input samples, output samples and the filter coefficients respectively. All experiments are conducted for a mission time of 1 month. Since SEUs can cause both SCUs or DCUs, for the combined model, it was assumed that 99\% ($\alpha_{SCU} = 0.99 $) of the SEUs will cause SCUs and 1\% ($\alpha_{DCU} = 0.01 $) of them will cause DCUs (with another added assumption that $\beta = \beta1$). Since the model is parametric, any other values for scaling the SCU and DCU rates can be used. Also, we use $\lambda_{voter}$ = 0 for the first few experiments, and then introduce a non-zero failure rate for the voters (using Equation \ref{domain}) to evaluate its impact on our models. We analyze four design options (as shown in \tablename~\ref{tab:model_stat}) using our methodology. We use the PRISM model checker version 4.1 to analyze the reliability and availability properties for each of them.\\

\noindent Before analyzing the model quantitatively, we verify the following LTL style properties (Recall Remark~\ref{sloop}) to check the correctness of the model:  \\

\noindent \emph{Correctness Property:} \texttt{filter(forall, P>0~[X~oper])} - ``From any reachable state, it is possible to reach the \emph{oper} state in the next step with a probability greater than 0 ".\\

Note that, as mentioned earlier in the preliminary section, blind scrubbing periodically reconfigures the FPGA, which does not require any fault detection. This sets the requirement (specified as the property above) that the system should be repaired irrespective of its failures, i.e. will be scrubbed even in the \emph{oper} state, which justifies the self-loop in our model. While verifying, PRISM returned \texttt{true}, which means that the correctness property hold in our model.

\begin{figure}[!t]
	\centering
	\includegraphics[width = \textwidth]{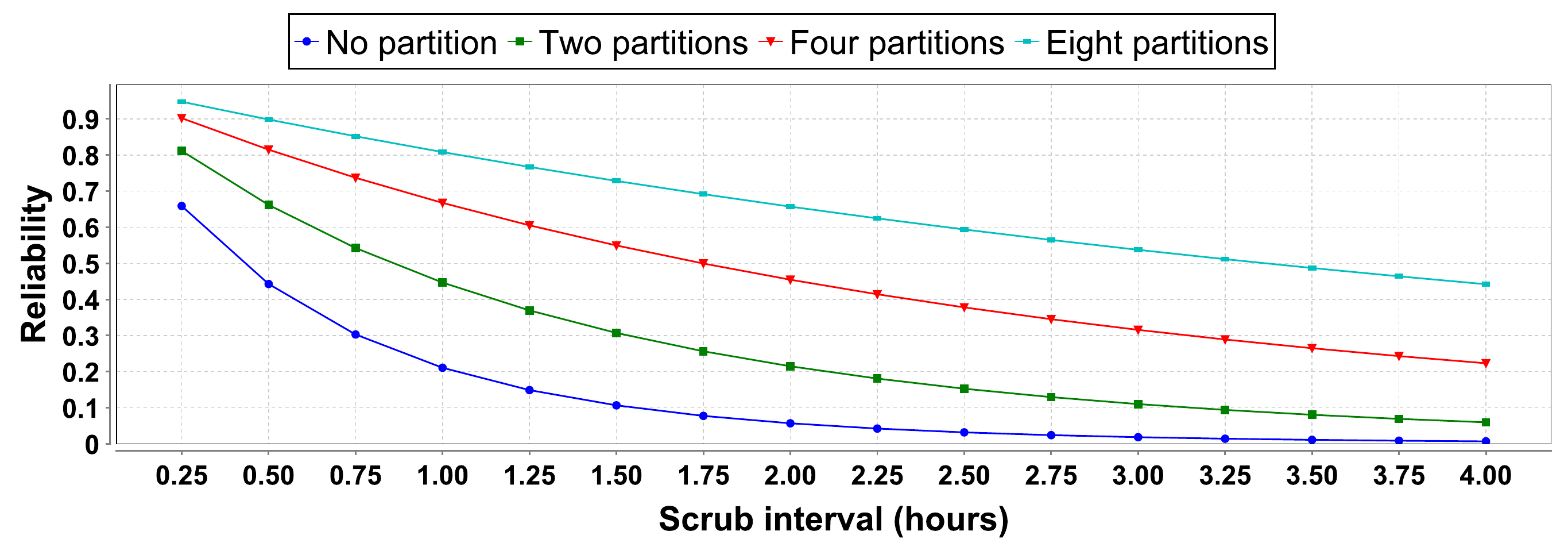}
	\caption{SCU Reliability}\label{Fig:rel}
\end{figure}
\subsection{Analysis}
\figurename~\ref{Fig:rel} shows the relationship between the reliability and number of partitions in the design for different scrub intervals using the SCU model. Reliability of a system (or component) is defined as the probability that the system performs correctly for a given period of time, from zero ($t_0$) to $t_1$, given that the system or the component was functioning correctly at $t_0$. In PRISM, this property can be formalized in CSL as \texttt{P=?[G[0,T] operational], T = 1 month}, and we evaluate this property for different scrub intervals starting from 15 minutes up to 4 hours. For all the design options with different scrub intervals, the reliability decreases when the scrub interval increases. However, designs with more partitions show significant improvements in reliability, even with the same scrub interval. For example, if the scrubbing interval is 15 minutes, the design with no partition has a reliability of 0.65 only. In contrast, the design with two, four and eight partitions has a reliability of 0.81, 0.90 and 0.94 respectively. TMR increases the area and power consumption by a factor of 300\% as a result of replications. More frequent scrub in such cases will consume more power that might not be appropriate for most space applications. For such circumstances, increasing the number of partitions can offer a good solution instead of a more frequent scrub strategy. For example, if the designer is targeting a reliability higher than 0.80, and if the design has no partition (or fewer partitions), then the designer may consider adopting a more frequent scrubbing strategy (less than an hour, in order of seconds or milliseconds) to meet some requirements. Instead of adding such power burden on the system, the designer may adopt TMR with 2, 4 or 8 partitions, which will require scrubbing once per 15 minutes (thus reducing power consumption) and will also meet the requirement. Note that the design option with eight partitions provides a reliability of 0.8 even for a delayed scrub of 1 hour. Using this approach a designer can determine the number of partitions required to meet the design requirements for a given scrub rate or vice versa.

\begin{figure}[!t]
\centering
\includegraphics[width = \textwidth]{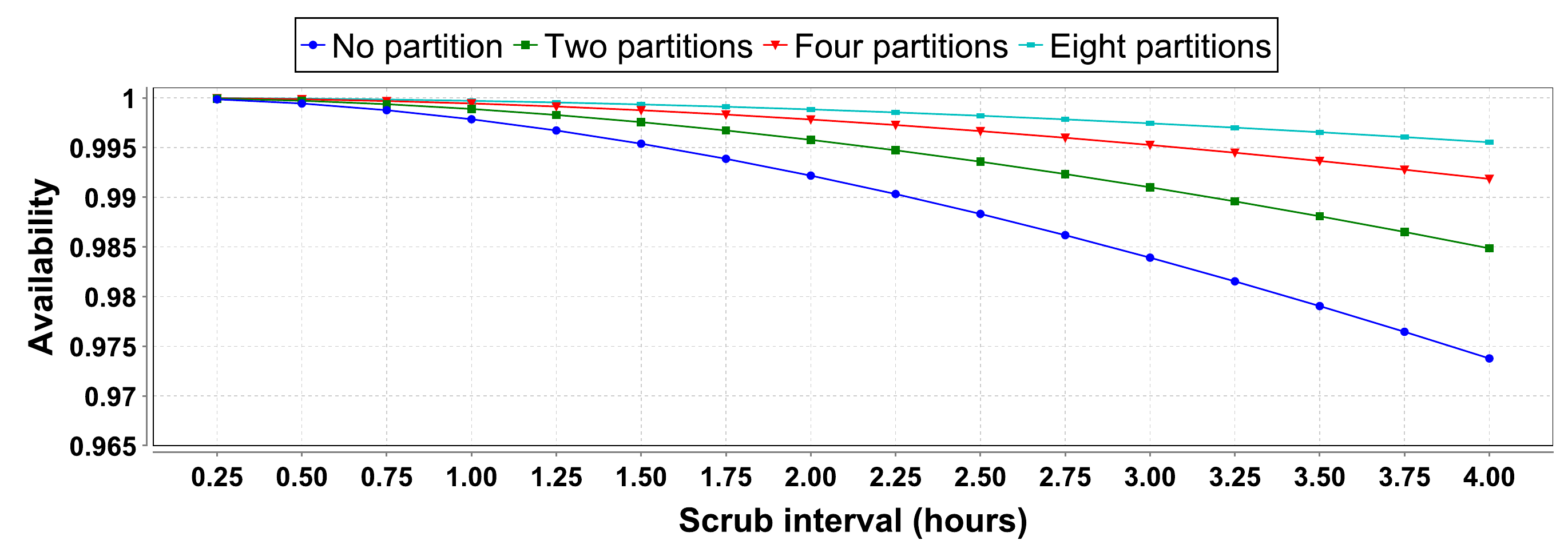}
\caption{SCU Availability}\label{Fig:avl}
\end{figure}

Availability is defined as the ratio of time the system or component operates correctly (system uptime) to its entire mission time. Using the SCU model, \figurename~\ref{Fig:avl} shows the availability of the design for different scrub intervals and a different number of partitions. In PRISM, this property can be formalized in CSL as  \texttt{R\{"up\_time"\}=? [C<=T]/T, T = 1 month}. The design with no partition offers an availability of 4 nines (0.9999) for the scrub interval of 15 minutes which drops up to 1 nine (0.97) with increased scrub interval of 4 hours. Compared to this, all the other options with TMR partitioning offer improved availability. For instance, for a scrub interval of three hours, the design with no partition offers only 98\% availability, whereas the rest of the design options with partitioning offers availabilities of more than 99\%. Most of the communication satellites target more than 99\% availability. In such cases, if the power constraint restricts the designer not to increase the scrub interval, then increasing the number of partitions may offer a solution.

%

\begin{figure}[!t]
	\centering
	\includegraphics[width = \textwidth]{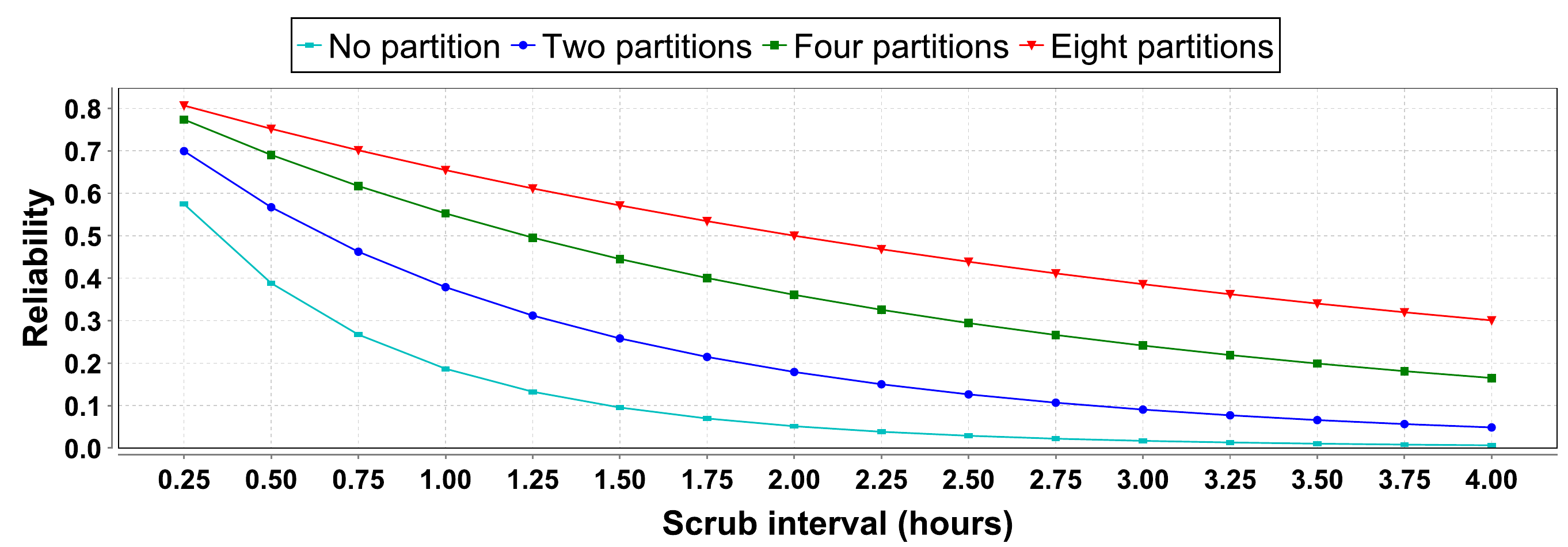}
	\caption{Combined Reliability}\label{Fig:relDBU}
\end{figure}

\begin{figure}[!t]
	\centering
	\includegraphics[width = \textwidth]{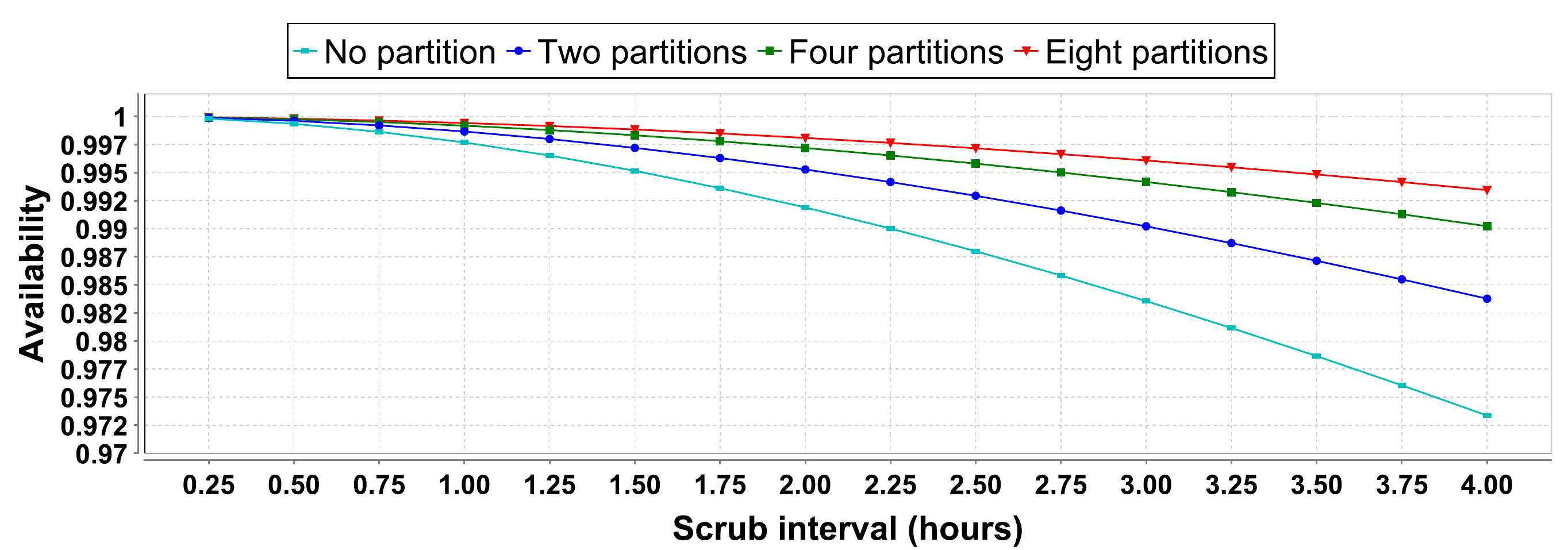}
	\caption{Combined Availability}\label{Fig:AvlDBU}
\end{figure}

For the second part of our analysis, we evaluated the same reliability and availability properties using the combined model for all the design options with two, four and eight partitions. The results are shown in \figurename~\ref{Fig:relDBU} and \figurename~\ref{Fig:AvlDBU}. We observe that even for the combined model, with increasing number of partitions, the reliability and availability of the design increases. Interestingly, increasing the number of partitions has a small effect on the value of the availability when the design employs frequent scrubbing, however, it should be noted that approaching 1 even by a small amount can be extremely difficult, and the improvement of availability is not well reflected on a linear scale. In contrast, increasing the number of partitions improves the availability dramatically for the cases where the scrub interval in comparatively longer.  From this, we can conclude that our observation for the SCU only model also holds for the combined model.



\subsection{Observation}A major observation from these analysis is that when the scrub interval is smaller (frequent scrub), the number of partitions plays a major role increasing the reliability of a system. However, even for a delayed scrub, the improvement is noticeable enough. In other words, the graphs show a trend that to meet the designer's reliability goal, if the number of partitions (which means smaller domains) is increased, less frequent scrub will be required. In contrast, fewer partitions (larger domain size) will require more frequent scrubs to meet a target reliability requirement. For the availability, the number of partitions plays a significant role for longer scrub intervals. For frequent scrub intervals, the number of partitions increases the availability to a minimal level, but for longer scrub intervals, the improvement of availability with the number of partitions is quite significant. Such early analysis on the high-level design description will allow a designer to perform the analysis before the actual implementation of the system considering the design constraints such as power. Using such methodology a designer can find better trade-offs between the number of partitions and the required scrub interval that will meet the design requirements, and also reduce the design effort, time and cost.

\subsection{Impact of the voter failure}
\begin{figure}[!t]
	\centering
	\includegraphics[width = \textwidth]{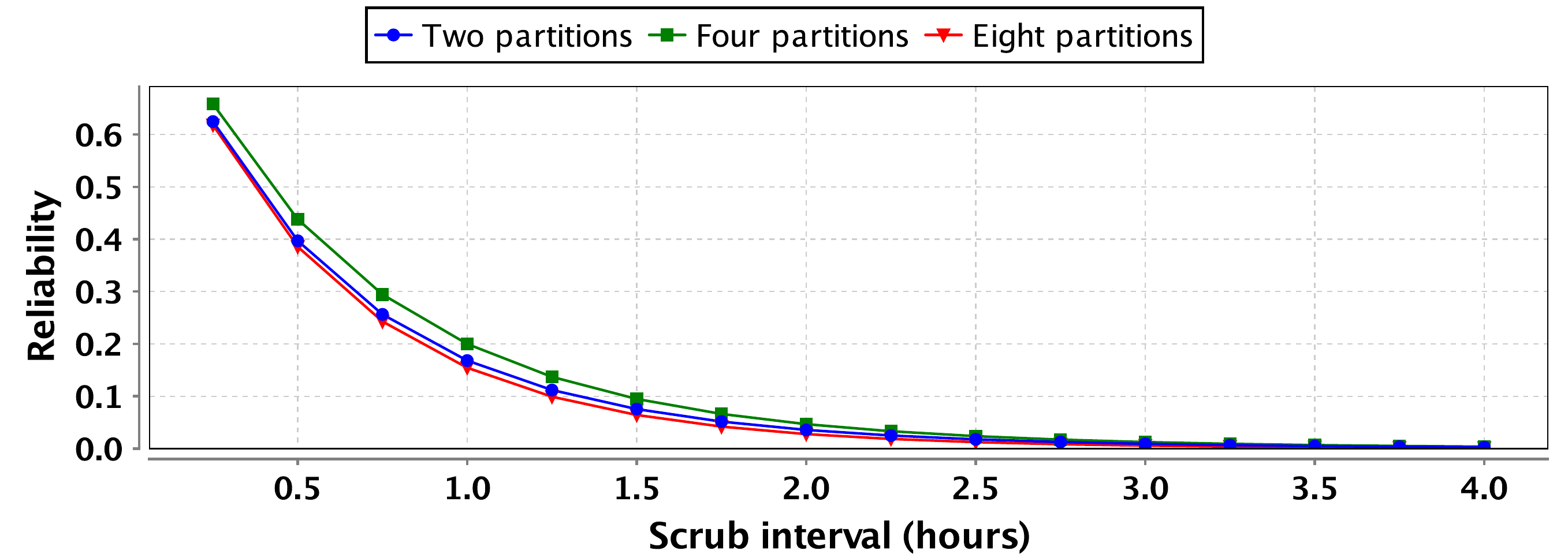}
	\caption{Impact of voter failure on the SCU model}\label{Fig:AvlvfSCU}
\end{figure}

\begin{figure}[!t]
	\centering
	\includegraphics[width = \textwidth]{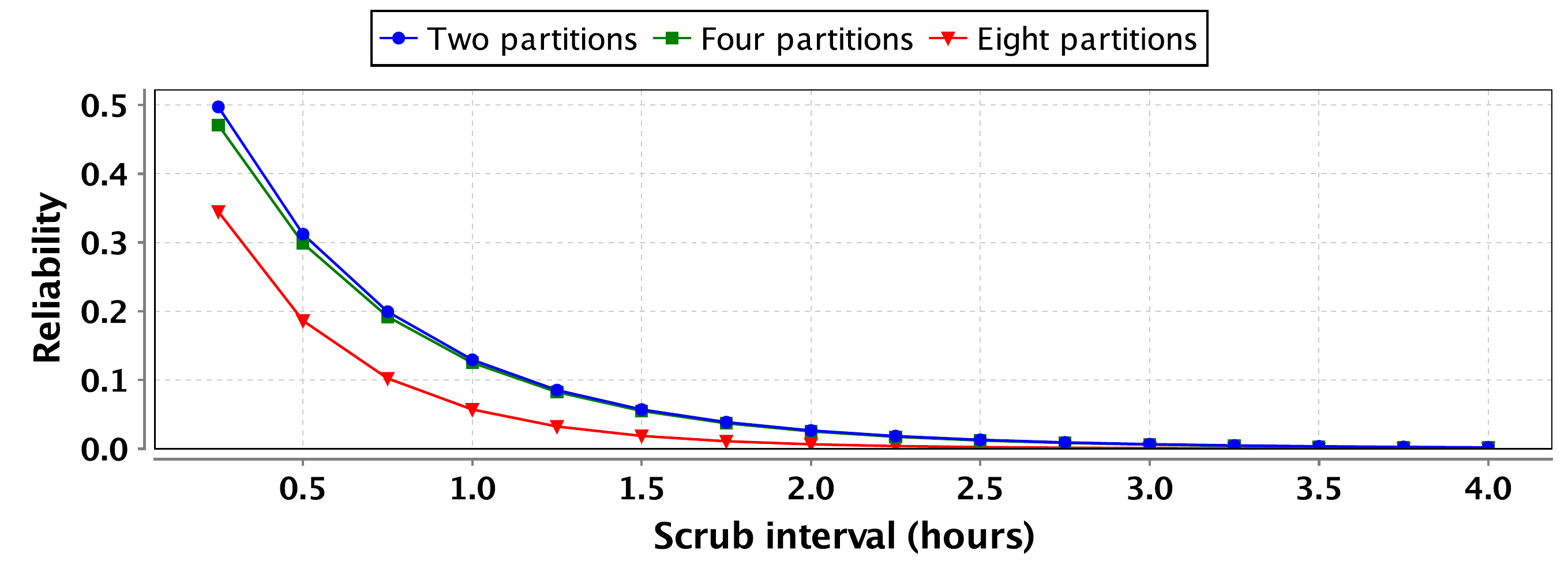}
	\caption{Impact of voter failure on the combined model}\label{Fig:AvlvfDBU}
\end{figure}
So far, we evaluated the proposed models assuming that TMR voters are not prone to failure. To illustrate the capability of our model in this section we introduce a non-zero failure rate ($\lambda_{voter} = 5 \times 10^{-3}$ per hour) for voters and evaluate both the SCU and combined model for two, four and eight partitions. The obtained results are shown in \figurename~\ref{Fig:AvlvfSCU} and \figurename~\ref{Fig:AvlvfDBU}. Interestingly, this time we observe a trade-off, and the obtained results clearly contrast the results obtained earlier with non-zero voter failure rate. For the SCU model, TMR with four partitions shows the best reliability compared to the options with two and eight partitions, unless the scrub interval is very long (for example if the scrub interval is more than 3 hours, then all the three TMR options offer a similar reliability). However, for the combined model, the reliability of the TMR with two partitions beats the other TMR versions. This finding is very interesting, since it shows a reversal of trends compared to what we observed when the TMR voters were assumed to be fault-free (in which more partition was always better). Also, when the scrub interval is more than an hour for the combined model, the reliabilities of TMR with two and four partitions become very close to each other. It means that if a designer needs to adopt a very long scrub interval, then TMR with two or four partitions, can both be a good choice for the design. In contrast, if it is possible to adopt a faster scrubbing (interval less than an hour), then TMR with two partitions is the proper choice. 

Design of a fault-tolerant voter has been an important and active research area for many years \cite{voterTMR1, voterTMR2}. Based on the choice of the voter to implement with the TMR partitions, it is straightforward to use our proposed approach for evaluating the relationship between the scrub interval and the number of TMR partitions knowing the specific choice of a voter and associated failure rate by changing only the parameters of the model.  

\subsection{Scalability} \label{ssec:scale}
As shown in \tablename~\ref{tab:model_stat}, when increasing number of partitions, the number of states also increase. The relationship between the number of partitions $N$ and the number of states $S_{total}$ can be expressed as $S_{total} = 3^N$, where $N\geq1$.  PRISM model checker includes multiple model checking engines, many of which are based on symbolic implementations (using binary decision diagrams and their extensions). These engines enable the probabilistic verification of models of up to $10^{10}$ states (on average, PRISM handles models with up to $10^7-10^8$ states). So based on this fact, with the proposed approach we can model up to 16 partitions and this is due to the limitation of the PRISM tool.

It is worth mentioning that in order to analyze more larger of partitions, it is possible to reduce the state space of the probabilistic models prior to verification. A variety of techniques has been recently developed including symmetry reduction, bisimulation minimisation and abstraction. Among these bisimulation (also known as lumping)\cite{buchholz1994exact, dehnert2013smt} is of particular interest for our future work, since it preserves widely used probabilistic temporal logics.  PRISM also features a variety of advanced techniques such as abstraction refinement and symmetry reduction. PRISM also supports approximate/statistical model checking through a discrete event simulation engine. So considering the capability of the PRISM model checker, it is also possible to analyze systems with larger number of partitions using our methodology.

\section{Conclusion and Future Works}

We presented the formal modeling and analysis of single-cell and multiple-cell upsets using a methodology based on the probabilistic model checking technique. This methodology aims to analyze the relationship between the number of TMR partitions, scrub interval and mission time. Increasing the number of TMR partitions allows to reduce the frequency of scrubbing, which results in less energy consumption. However, based on the voter failure rate, it is possible to find the optimal number of partitions. Using the proposed methodology, designers can assess the number of partitions, or the scrub frequency required to meet the design requirement at early design stages. To demonstrate our approach, we have shown the results of our analysis for a 64-tap FIR filter case study. The results showed how the increased number of partitions enable less frequent scrubs and vice-versa, and also we were able to find the optimal number of partitions for both: the SCU and the combined model. Such an early analysis will add more confidence to the design and may reduce the overall design cost, time, and effort. However, since the PRISM modeling is not automated, this restricts us to do so. In the future, we will work to overcome this limitation. It is also worth mentioning that with the decrease of the transistor size, three or more bit upsets are also not uncommon these days. This will be addressed in our future work. Another interesting future work for us will be to include the partial reconfiguration (read-back scrubbing) in our model and to explore the effect of unreliable voters in the design partitions.


\section*{Acknowledgments}
This research work is a part of the AVIO-403 project financially supported by the Consortium for Research and Innovation in Aerospace in Quebec (CRIAQ), Fonds de Recherche du Qu\'ebec - Nature et Technologies (FRQNT) and the Natural Sciences and Engineering Research Council of Canada (NSERC). The authors would also like to thank Bombardier Aerospace, MDA Space Missions and the Canadian Space Agency (CSA) for their technical guidance and financial support.


\bibliographystyle{elsarticle-num}
\bibliography{myBib}

\begin{thebibliography}{10}
\expandafter\ifx\csname url\endcsname\relax
  \def\url#1{\texttt{#1}}\fi
\expandafter\ifx\csname urlprefix\endcsname\relax\def\urlprefix{URL }\fi
\expandafter\ifx\csname href\endcsname\relax
  \def\href#1#2{#2} \def\path#1{#1}\fi

\bibitem{XilinxRosetta}
A.~Lesea, Continuing experiments of atmospheric neutron effects on deep
  submicron integrated circuits ({WP286} v1.1) (October 2011).

\bibitem{XilinxTMR}
C.~Carmichael, Triple module redundancy design techniques for virtex {FPGA}s
  ({XAPP}197 v1.0.1), {X}ilinx corporation, 2006.

\bibitem{NASAScrub}
P.~Adell, G.~Allen, G.~Swift, S.~McClure, Assessing and mitigating radiation
  effects in {X}ilinx {SRAM} {FPGA}s, in: Radiation and Its Effects on
  Components and Systems (RADECS), 2008 European Conference on, 2008, pp.
  418--424.

\bibitem{Related-2TMR}
B.~H. Pratt, M.~P. Caffrey, D.~Gibelyou, P.~S. Graham, K.~Morgan, M.~J.
  Wirthlin, {TMR} with more frequent voting for improved {FPGA} reliability.,
  in: ERSA, 2008, pp. 153--158.

\bibitem{Related-3TMRoptimal}
F.~L. Kastensmidt, L.~Sterpone, L.~Carro, M.~S. Reorda, On the optimal design
  of triple modular redundancy logic for {SRAM}-based {FPGA}s, in: Proceedings
  of the conference on Design, Automation and Test in Europe-Volume 2, IEEE
  Computer Society, 2005, pp. 1290--1295.

\bibitem{PRISM:caseStudy}
M.~Kwiatkowska, G.~Norman, D.~Parker, Controller dependability analysis by
  probabilistic model checking, Control Engineering Practice 15~(11) (2006)
  1427--1434.

\bibitem{formal1}
S.~D. Nelson, C.~Pecheur, Formal verification for a next-generation space
  shuttle, in: International Workshop on Formal Approaches to Agent-Based
  Systems, Springer, 2002, pp. 53--67.

\bibitem{formal2}
C.~Rouff, A.~Vanderbilt, W.~Truskowski, J.~Rash, M.~Hinchey, Verification of
  nasa emergent systems, in: Engineering Complex Computer Systems, 2004.
  Proceedings. Ninth IEEE International Conference on, IEEE, 2004, pp.
  231--238.

\bibitem{formal3}
S.~Kothari, Software engineering research lab to airplanes, orion and beyond,
  in: Software Engineering Research and Industrial Practice (SER\&IP), 2016
  IEEE/ACM 3rd International Workshop on, IEEE, 2016, pp. 3--9.

\bibitem{khaza}
K.~A. Hoque, O.~Ait~Mohamed, Y.~Savaria, C.~Thibeault, Early analysis of soft
  error effects for aerospace applications using probabilistic model checking,
  in: C.~Artho, P.~C. Ölveczky (Eds.), Formal Techniques for Safety-Critical
  Systems, Vol. 419 of Communications in Computer and Information Science,
  Springer International Publishing, 2014, pp. 54--70.

\bibitem{hoque2017formal}
K.~A. Hoque, O.~Ait~Mohamed, Y.~Savaria, {Formal analysis of SEU mitigation for
  early dependability and performability analysis of FPGA-based space
  applications}, Journal of Applied Logic 25~(Supplement C) (2017) 47 -- 68.
\newblock \href {http://dx.doi.org/https://doi.org/10.1016/j.jal.2017.03.001}
  {\path{doi:https://doi.org/10.1016/j.jal.2017.03.001}}.

\bibitem{miner2000case}
P.~S. Miner, V.~A. Carre{\~n}o, M.~Malekpour, W.~Torres, A case-study
  application of rtca do-254: design assurance guidance for airborne electronic
  hardware, in: Digital Avionics Systems Conference, 2000. Proceedings. DASC.
  The 19th, Vol.~1, IEEE, 2000, pp. 1A1--1.

\bibitem{FAAFM}
{Brian Butka}, {Advanced Verification Methods for Safety-Critical Airborne
  Electronic Hardware},
  \url{https://www.faa.gov/aircraft/air_cert/design_approvals/air_software/media/TC-14-41.pdf},
  {Federal Aviation Administration Report No: DOT/FAA/TC-14/41} (2015).

\bibitem{Baier99approximatesymbolic}
C.~Baier, J.-P. Katoen, H.~Hermanns, Approximate symbolic model checking of
  continuous-time markov chains (extended abstract) (1999).

\bibitem{PMC}
{PRISM website}, http://www.prismmodelchecker.org.

\bibitem{LibraryBasedSER}
C.~Thibeault, Y.~Hariri, S.~R. Hasan, C.~Hobeika, Y.~Savaria, Y.~Audet, F.~Z.
  Tazi, A library-based early soft error sensitivity analysis technique for
  {SRAM}-based {FPGA} design, J. Electronic Testing 29~(4) (2013) 457--471.

\bibitem{fault1}
W.~Mansour, R.~Velazco, {SEU} fault-injection in {VHDL}-based processors: A
  case study, J. Electronic Testing 29~(1) (2013) 87--94.

\bibitem{SEUFault}
P.~Kenterlis, N.~Kranitis, A.~M. Paschalis, D.~Gizopoulos, M.~Psarakis, A
  low-cost {SEU} fault emulation platform for {SRAM}-based {FPGA}s, in: IOLTS,
  2006, pp. 235--241.

\bibitem{Khaza_MEMOCODE2014}
K.~A. Hoque, O.~A. Mohamed, Y.~Savaria, C.~Thibeault, {Probabilistic Model
  Checking Based DAL Analysis to Optimize a Combined TMR-Blind-Scrubbing
  Mitigation Technique for FPGA-Based Aerospace Applications}, in:
  International Conference on Formal Methods and Models for Co-Design,
  ACM-IEEE, 2014.

\bibitem{ret1}
P.~Hazucha, C.~Svensson, Impact of cmos technology scaling on the atmospheric
  neutron soft error rate, Nuclear Science, IEEE Transactions on 47~(6) (2000)
  2586--2594.
\newblock \href {http://dx.doi.org/10.1109/23.903813}
  {\path{doi:10.1109/23.903813}}.

\bibitem{ret5}
A.~Lesea, S.~Drimer, J.~Fabula, C.~Carmichael, P.~Alfke, {The rosetta
  experiment: atmospheric soft error rate testing in differing technology
  FPGAs}, Device and Materials Reliability, IEEE Transactions on 5~(3) (2005)
  317--328.
\newblock \href {http://dx.doi.org/10.1109/TDMR.2005.854207}
  {\path{doi:10.1109/TDMR.2005.854207}}.

\bibitem{sterpone2007experimental}
L.~Sterpone, M.~Violante, R.~H. Sorensen, D.~Merodio, F.~Sturesson, R.~Weigand,
  S.~Mattsson, Experimental validation of a tool for predicting the effects of
  soft errors in {SRAM-based FPGAs}, Nuclear Science, IEEE Transactions on
  54~(6) (2007) 2576--2583.

\bibitem{rel-TMR-2015}
L.~Tambara, F.~Almeida, P.~Rech, F.~Kastensmidt, G.~Bruni, C.~Frost,
  \href{http://dx.doi.org/10.1007/978-3-319-16214-0_28}{{Measuring Failure
  Probability of Coarse and Fine Grain TMR Schemes in SRAM-based FPGAs Under
  Neutron-Induced Effects}}, in: Applied Reconfigurable Computing, Vol. 9040 of
  Lecture Notes in Computer Science, Springer International Publishing, 2015,
  pp. 331--338.
\newline\urlprefix\url{http://dx.doi.org/10.1007/978-3-319-16214-0_28}

\bibitem{Rel-TMR-AnalyticalModel-Zhong}
Z.-M. Wang, L.-L. Ding, Z.-B. Yao, H.-X. Guo, H.~Zhou, M.~Lv, {The reliability
  and availability analysis of SEU mitigation techniques in SRAM-based FPGAs},
  in: Radiation and Its Effects on Components and Systems (RADECS), 2009
  European Conference on, 2009, pp. 497--503.
\newblock \href {http://dx.doi.org/10.1109/RADECS.2009.5994702}
  {\path{doi:10.1109/RADECS.2009.5994702}}.

\bibitem{AHS2013}
L.~Sterpone, A.~Ullah, {On the optimal reconfiguration times for TMR circuits
  on SRAM based FPGAs}, in: Adaptive Hardware and Systems (AHS), 2013 NASA/ESA
  Conference on, IEEE, 2013, pp. 9--14.

\bibitem{rel-TMR-FineGrain}
B.~Pratt, M.~Caffrey, J.~Carroll, P.~Graham, K.~Morgan, M.~Wirthlin,
  {Fine-Grain SEU Mitigation for FPGAs Using Partial TMR}, Nuclear Science,
  IEEE Transactions on 55~(4) (2008) 2274--2280.
\newblock \href {http://dx.doi.org/10.1109/TNS.2008.2000852}
  {\path{doi:10.1109/TNS.2008.2000852}}.

\bibitem{remAfter}
F.~Ling, M.~Chunyan, C.~Zhuo, L.~Guoqiang, Evaluation of redundancy based
  system: A model checking approach, SCIENCE CHINA Information Sciences
  null~(null).

\bibitem{Clarke86automaticverification}
E.~M. Clarke, E.~A. Emerson, A.~P. Sistla, Automatic verification of
  finite-state concurrent systems using temporal logic specifications, ACM
  Transactions on Programming Languages and Systems 8 (1986) 244--263.

\bibitem{pnueli1977temporal}
A.~Pnueli, \href{http://dx.doi.org/10.1109/SFCS.1977.32}{The temporal logic of
  programs}, in: Proceedings of the 18th Annual Symposium on Foundations of
  Computer Science, SFCS '77, IEEE Computer Society, Washington, DC, USA, 1977,
  pp. 46--57.
\newblock \href {http://dx.doi.org/10.1109/SFCS.1977.32}
  {\path{doi:10.1109/SFCS.1977.32}}.
\newline\urlprefix\url{http://dx.doi.org/10.1109/SFCS.1977.32}

\bibitem{GAUT}
{GAUT}: A high-level synthesis tool for {DSP} applications, in: P.~Coussy,
  A.~Morawiec (Eds.), High-Level Synthesis, 2008.

\bibitem{hoque2015towards}
K.~A. Hoque, O.~A. Mohamed, Y.~Savaria, Towards an accurate reliability,
  availability and maintainability analysis approach for satellite systems
  based on probabilistic model checking, in: Design, Automation, and Test in
  Europe, IEEE, 2015.

\bibitem{XilinxScrub}
A.~S. C.~Carmichael, M.~Caffrey, Correcting {S}ingle-{E}vent {U}psets through
  {V}irtex {P}artial {C}onfiguration ({XAPP}216 v1.0), {X}ilinx corporation,
  2010.

\bibitem{intvsExtScrub}
M.~Berg, C.~Poivey, D.~Petrick, D.~Espinosa, A.~Lesea, K.~LaBel, M.~Friendlich,
  H.~Kim, A.~Phan, Effectiveness of internal versus external {SEU} scrubbing
  mitigation strategies in a {X}ilinx {FPGA}: Design, test, and analysis,
  Nuclear Science, IEEE Transactions on 55~(4) (2008) 2259--2266.

\bibitem{concurrency}
H.~Hermanns, L.~Zhang, From concurrency models to numbers: Performancd and
  dependability, in: Nato Science for Peace and Security Series, IOS Press,
  2011, pp. 182--210.

\bibitem{PMC2prism}
M.~Kwiatkowska, G.~Norman, D.~Parker, {PRISM} 4.0: Verification of
  probabilistic real-time systems, in: Computer aided verification, Springer,
  2011, pp. 585--591.

\bibitem{FIRdsp}
S.~Parkes, {DSP (Demanding Space-based Processing!): the Path Behind and the
  Road Ahead, DSP’98}, in: 6th International Workshop on Digital Signal
  Processing Techniques for Space Applications, ESTEC, Noordwijk, The
  Netherlands, 1998, pp. 23--25.

\bibitem{FIR-3}
T.~M. Braun, Satellite Communications payload and system, John Wiley \& Sons,
  2012.

\bibitem{voterTMR1}
P.~Balasubramanian, K.~Prasad, N.~E. Mastorakis, {A fault tolerance improved
  majority voter for TMR system architectures}, arXiv preprint
  arXiv:1605.03771.

\bibitem{voterTMR2}
R.~V. Kshirsagar, R.~M. Patrikar, Design of a novel fault-tolerant voter
  circuit for {TMR} implementation to improve reliability in digital circuits,
  Microelectronics Reliability 49~(12) (2009) 1573--1577.

\bibitem{buchholz1994exact}
P.~Buchholz, Exact and ordinary lumpability in finite markov chains, Journal of
  applied probability (1994) 59--75.

\bibitem{dehnert2013smt}
C.~Dehnert, J.-P. Katoen, D.~Parker, {SMT-based bisimulation minimisation of
  Markov models}, in: International Workshop on Verification, Model Checking,
  and Abstract Interpretation, Springer, 2013, pp. 28--47.

\end{thebibliography}


%
%

\end{document}